\documentclass[prd,showpacs,showkeys,superscriptaddress]{revtex4-2}
\usepackage{amsmath}
\usepackage{mathrsfs}
\usepackage{lipsum}
\usepackage{amssymb}
\usepackage{latexsym}
\usepackage{graphics}
\usepackage{subfigure}
\usepackage[colorlinks=true, linkcolor=blue, citecolor=red, urlcolor=magenta]{hyperref}
\usepackage{amsfonts}
\usepackage{graphicx}
\usepackage{calrsfs}
\usepackage{epsfig}
\usepackage{comment}

\def\P{{\cal{P}}}

\def\eqref#1{{(\ref{#1})}}

\def\be{\begin{equation}}
\def\ee{\end{equation}}
\def\beq{\begin{eqnarray}}
\def\eeq{\end{eqnarray}}

\usepackage{orcidlink}

\begin{document}

\title{Shadow of a Noncommutative Thin-Shell Gravastar}

\author{M. A. Anacleto~\orcidlink{0000-0003-4625-7322}}
\email{anacleto@df.ufcg.edu.br}
\affiliation{Departamento de F\'{\i}sica, Universidade Federal de Campina Grande
Caixa Postal 10071, 58429-900 Campina Grande, Para\'{\i}ba, Brazil}
\affiliation{Unidade Acad\^emica de Matem\'atica, Universidade Federal de Campina Grande
\\
58429-900 Campina Grande, Para\'{\i}ba, Brazil}
\author{A. T. N. Silva~\orcidlink{0009-0007-0067-600X}}
\email{andersont.nsilva@gmail.com}
\affiliation{Departamento de F\'{\i}sica, Universidade Federal de Campina Grande
Caixa Postal 10071, 58429-900 Campina Grande, Para\'{\i}ba, Brazil}
\author{L. Casarini~\orcidlink{0000-0002-0869-9405}}
\email{lcasarini@academico.ufs.br}
\affiliation{Departamento de F\'isica, Universidade Federal de Sergipe
49100-000 Aracaju, Sergipe, Brazil}

\begin{abstract}
\noindent One of the main challenges in astronomy is the direct observation of black holes. However, distinguishing black holes from ultra-compact horizonless objects through photon observations can be difficult, since the latter may also possess unstable circular photon orbits. An example of an  Ultra-Compact Object is a Gravastar (gravitational vacuum star), initially proposed by Mazur and Mottola. For definition purposes, we construct a spherically symmetric thin-shell gravastar model within a noncommutative model, in which these effects are integrated by a Lorentzian distribution of the energy density with a minimum width $\sqrt{\theta}$. The model is constructed using the cut-and-paste technique to connect a nonsingular de Sitter interior to a noncommutative Schwarzschild exterior, satisfying the Israel junction conditions on the interface hypersurface $\Sigma$. We examine the stability of the model through its energy conditions, highlighting the influence of the noncommutative parameter on its behavior. {Model stability was also investigated using the parameter $\eta$, derived from the linearized stability analysis.} We also show that, owing to noncommutativity, the photon trajectories and deflection angles change as the noncommutative parameter increases. Our proposed gravastar model, with noncommutative geometry on its exterior, can be considered a stable and viable alternative to the charged black hole in the context of this type of gravity.


\end{abstract}

\maketitle

\section{Introduction}
In recent years, cosmology and astrophysics have played a central role in the search for essential answers about the nature of the universe. Advances in observations and the creation of new theoretical frameworks have expanded our understanding of cosmic structures and the astrophysical processes that shape the universe, while also revealing conflicts with traditional models and enabling the introduction of concepts still in their consolidation phase. For this reason, compact objects stand out as privileged laboratories, as they meet the extreme conditions possible for testing hypotheses in an environment of high densities and strong gravitational fields. Black holes, for instance, are supported by compelling observational evidence, including the imaging of shadows at the center of the galaxy M87~\cite{EventHorizonTelescope:2019dse,EventHorizonTelescope:2021bee, Banerjee:2019nnj} and of the nearby Sgr $A^{*}$~\cite{EventHorizonTelescope:2022wkp, Banerjee:2022jog}, the radiation emitted by accretion disks in active galactic nuclei~\cite{Czerny:2022xfj}, as well as the direct detection of gravitational waves from merging binary black hole systems by LIGO/Virgo~\cite{KAGRA:2021duu}. However, although such evidence provides a coherent description, it raises conceptual problems that have not yet been fully resolved by the scientific community: the existence of singularities and the information loss paradox.

Given these problems, scientists are seeking alternative models that produce the observational properties of a black hole while avoiding the aforementioned obstacles. Aiming to address these issues, Mazur and Mottola \cite{Mazur} proposed that a phase transition could prevent complete stellar collapse, thereby giving rise to the gravitational vacuum star model, known as Gravastar, an object that has neither a singularity nor an event horizon. This model consists of three regions: an inner dark energy core (described by de Sitter geometry) surrounded by a thin layer of exotic matter and an asymptotically flat exterior (originally Schwarzschild), connected through the cut-and-paste technique and the Israel junction conditions \cite{Israel, Lobo:2015lbc}. The lack of a horizon prevents the information paradox, while the negative energy in the core stabilizes the central singularity \cite{Lobo:2003xd}. These types of stellar objects help us understand the role of dark energy in the rapid expansion of the Universe, as well as helping to explain why some galaxies have more or less dark matter.

The gravastar concept originated from the observation that a condensed-matter system may undergo a phase transition under suitable conditions. Given the significant quantum mechanical effects near the event horizon, the behavior of collapsing dust particles is predicted to resemble that of a quantum system. The significant quantum mechanical effects near the event horizon suggest that the behavior of collapsing dust particles resembles that of a multi-interacting quantum system. It is known that at very low temperatures, a collection of bosonic atoms or molecules can condense into the Bose-Einstein state. Mazur and Mottola investigated this condensation and applied it to celestial bodies undergoing gravitational collapse, developing the concept of a hypothetical compact, cold, and dark object. They named it a gravitational vacuum condensate star, or gravastar, to circumvent the issues of singularity and the event horizon. Thus, the model was seen as an alternative to the classical black hole. According to their model, gravastar has three distinct regions with different equations of state (EoS): an inner region filled with dark energy in an isotropic de Sitter vacuum ($p = -\rho$), a thin-shelled intermediate layer composed of rigid fluid matter ($p = \rho$), and an empty outer region in which the Schwarzschild geometry ($p=\rho = 0$) adequately describes this condition~\cite{Pani:2009hk,Uchikata:2016qku,Kubo:2016ada,Pani:2015tga,Sakai:2014pga,Pani:2012zz,MartinMoruno:2011rm,Pani:2010em,Pani:2009ss,Rocha:2014gza,Rocha:2014jma,Matos}.

Recent research on the brightness of distant type Ia supernovae~\cite{SupernovaSearchTeam:1998fmf, SupernovaCosmologyProject:1998vns, Bahcall:1999xn, Planck:2018vyg} indicates that the expansion of the universe is occurring more accelerated than previously thought. This indicates that the cosmic pressure $p$ and the energy density $\rho$ must violate the strong energy condition, that is, $ \rho + 3p < 0$. The component known as ``dark energy'' is what makes it possible for this demand to be met at a specific stage of cosmic evolution \cite{Sahni:1999gb, Peebles:2002gy, Padmanabhan:2002ji}. The dark energy condition is set by several substances. The most recognized proposal considers a non-zero cosmological constant, which is equivalent to the fluid satisfying the equation of state \(p = - \rho\) \cite{Pradhan:2023wac}.
The phase transition is motivated by quantum mechanics, in the limit where the behavior of collapsing matter makes the effects of quantum gravity significant, reorganizing itself to leave the gravastar stable. In this context, it is proposed that the interior of the object is filled with a form of vacuum energy (of the de Sitter type), surrounded by a thin shell of ultrarelativistic matter that separates this internal region from the exterior Schwarzschild~\cite{Chirenti:2016hzd,Visser:2003ge,Mottola:2023jxl,Mottola:2025fhl}.

To characterize this scenario, we use an extension of the spacetime structure that efficiently integrates possible corrections from quantum gravity, known as noncommutative geometry. In this formalism, the spacetime coordinates obey a commutation relation of the type $[\hat{x}^{\mu}, \hat{x}^{\nu}]$ = $i\vartheta^{\mu\nu}$, where $\hat{x}$ and $i\vartheta^{\mu\nu}$ represent the coordinate operators and an antisymmetric tensor of dimension (length)$^2$. As a result, energy-matter distributions that were previously considered point-like (delta functions) become smoothed, typically using Gaussian distributions with a minimum width of $\sqrt{\theta}$~\cite{Nicolini:2005vd}. This change resolves discrepancies related to gravitational collapse by removing the singularity at the core of the compact object.
In addition to noncommutativity eliminates is characterized by a Lorentzian function distribution with a minimal~\cite{Anacleto:2019tdj,Zeng:2021dlj,Zeng:2022fdm,Anacleto:2020efy,Anacleto:2020zfh,Hu:2023eow,Anacleto:2022shk,Saleem:2023pyx,AraujoFilho:2024rss,Hamil:2024ppj,Wang:2024fiz,Jha:2022bpv,Jha:2023htn}, i.e., a smeared particle, instead of the Dirac-delta function distribution. A way of implementing the energy density of a static and spherically symmetric, smeared and particle-like gravitational source has been considered in the following form \cite{Campos:2021sff}:
$\rho_{\theta}(r)={M\sqrt{\theta}}/{\pi^{3/2}(r^2+\pi\theta)^{2}}$,
where the mass $M$ is diffused throughout a region of linear dimension due $\sqrt{\theta}$
to the uncertainty. 
In the literature, studies on gravastar have been carried out in several models, such as noncommutative geometries~\cite{BANERJEE2016,Lobo:2010uc,Das:2018fzc}, 
modified~\cite{Sinha:2024xdy,Yousaf:2019zcb,Shamir:2018qhq,Shamir:2020apc,BHAR2021} and  
rainbow gravity~\cite{Barzegar:2023ueo} (see also~\cite{Rosa:2024bqv,Khlopov:1985jw,Konoplich:1999,Khlopov:2000,Khlopov:2008qy, AraujoFilho:2025jcu, Heidari:2025iiv} for other models). 

The study of black hole ``shadows'' has been a broad field of research in recent decades~\cite{cunha2018shadows,mishra2019understanding,KONOPLYA20191,haroon2020shadow,bisnovatyi2018shadow, Hu:2020usx, Zhong:2021mty, Wang:2025fmz, Battista:2023iyu}. Most of these analyses are based on the dynamics of unstable circular orbits of photons. In this scenario, the so-called ``shadows'' do not represent direct images of the event horizon; they are, in fact, dark areas that result from the intense deflection of light from optical sources located in their vicinity.
However, it is crucial to understand that the existence of an unstable circular photon orbit is not a characteristic exclusive to black holes.  Ultra-Compact Objects (UCOs), such as Gravastars, can also possess this type of orbit. This implies that the simple observation of temporal brightenings produced by infalling gas is not, in itself, conclusive evidence of the existence of a black hole. Therefore, it is essential to investigate the possibility of other supercompact objects and their observational consequences \cite{Sakai:2014pga}.

{In~\cite{Campos:2021sff}, Campos et al. developed the noncommutative Schwarzschild metric with a Lorentzian distribution and analyzed null geodesics in the exterior region of a black hole. Övgün et al.~\cite{Ovgun:2017jzt} constructed the thin-shell formalism for gravastars in a noncommutative geometry, but without investigating photon behavior. Sakai et al.~\cite{Sakai:2014pga} examined the shadows of classical gravastars without considering noncommutative corrections.}{Wang et al.~\cite{Wang:2020emr} discovered and identified that the impact parameter of null geodesics is generally discontinuous along the wormhole.}
{Therefore, in this work, we combine these three elements for the first time. We show that noncommutativity modifies the critical impact parameter, alters the energy conditions on the shell in such a way that the parameter $\Theta$ can play a functional role analogous to that of the cosmological constant $\Lambda$, and we provide an estimate for the noncommutative energy scale based on geometric and astrophysical constraints.}

The following is how our paper is structured. In Sec.~\ref{sec1} where the mass function is specified and noncommutative geometry is considered. We implemented the effect of noncommutativity in the Schwarzschild black hole metric by a Lorentzian smeared mass distribution. The structural equations of noncommutative gravastars are constructed in Sec.~\ref{sec2}, where we identify surface tensions and discuss the corresponding conditions at the junction interface and their energy conditions. Null geodesic equations are derived in Sec.~\ref{sec3}, and solutions with unstable circular photon orbits are found. The pictures of optical sources behind the gravastar are obtained by numerically solving the null geodesic equations.The stability regions are shown in Sec.~\ref{STA}, where we can identify the ranges in which the system remains stable for certain parameter values. Finally, the Sec.~\ref{sec4} is devoted to concluding remarks.
\newpage

\section{Noncommutative geometry inspired Lorentzian Smeared Mass Distribution} 
\label{sec1}

We begin this section by constructing the exterior metric of a gravastar in the noncommutative background. 
For this purpose, we introduce noncommutativity by considering a Lorentzian mass density as follows~\cite{Anacleto:2019tdj,Campos:2021sff}.
\begin{equation}
\rho_{\theta}(r)=\frac{M\sqrt{\theta}}{\pi^{3/2}(r^2+\pi\theta)^{2}},
\label{MD1}
\end{equation}
{where $\theta$ is the noncommutative parameter of dimension $length^2$ and $M$ is the total mass diffused throughout the region of linear size $\sqrt{\theta}$.} So we can find the smeared mass distribution function~\cite{Anacleto:2019tdj,Campos:2021sff}
\begin{eqnarray}
\mathcal{M}_{\theta}&=&\int_0^r\rho_{\theta}(r)4\pi r^2 dr
=\frac{2M}{\pi}\left[\tan^{-1}\left( \frac{r}{\sqrt{\pi\theta}} \right) -\frac{r\sqrt{\pi\theta}}{\pi\theta + r^2}  \right],
\\
&=& M-\frac{4 M\sqrt{\theta}}{\sqrt{\pi}r} + {\cal O}(\theta^{3/2}). 
\end{eqnarray}
Therefore, the line element referring 
in the noncommutative background is now given by
\begin{equation}
ds^{2}=-f(r)dt^{2}+f(r)^{-1}dr^{2}+r^{2}d\Omega^{2},
\end{equation}
\\
with the metric function given by~\cite{Anacleto:2019tdj,Campos:2021sff}
\beq
f(r)=1 - \frac{2 \mathcal{M}_{\theta}}{r}=1 - \frac{2 M}{r} + \frac{8M\sqrt{\theta}}{\sqrt{\pi}r^{2}} + {\cal O}(\theta^{3/2}),
\label{FNC}
\eeq
where the additional term arises directly from the treatment of smeared mass distribution function. In the limit where $ \frac{\sqrt{\theta}}{r} \ll 1$, the asymptotic behavior of the classical Schwarzschild metric is recovered, that is, $\mathcal{M}_\theta \rightarrow M$. This behavior ensures that for large distances, spacetime recovers the usual gravitational solution, while in regions close to the origin, the noncommutative effect dominates the gravitational solution.{The construction presented in Eqs.~(\ref{MD1})-(\ref{FNC}), based on the Lorentzian smeared mass distribution and its corresponding asymptotic expansion, was originally developed and discussed in detail in Ref.~\cite{Anacleto:2019tdj}. Here, we adopt this established formulation as the exterior geometry for the thin-shell gravastar configuration.} The metric function in Eq.(\ref{FNC}) has also been applied to other gravitational models within  the noncommutative geometry framework~\cite{Filho:2024zxx, AraujoFilho:2025rwr,AraujoFilho:2025rzh}, and {corresponds to the asymptotic expansion of the Lorentzian noncommutative Schwarzschild solution reinforcing their relevance. Similar asymptotic expansions have been employed in studies of noncommutative gauge theories, Lorentzian smeared sources, and black hole phenomenology~\cite{Heidari:2023egu,AraujoFilho:2024mvz, Juric:2025kjl}}

{ Therefore, the geometry adopted throughout this work should be understood as an approximate solution valid within this domain.  As a result, the construction of the gravastar model, the junction conditions, the energy-condition analysis, the photon sphere determination, the critical impact parameter, and the stability analysis are thus all carried out consistently within this approximation scheme.}

\section{STRUCTURE EQUATIONS OF NONCOMMUTATIVE GRAVASTARS}\label{sec2}
 Initially, we examine two spacetime manifolds to construct gravastars. ${M_{+}}$ defines the exterior, while ${M_{-}}$ defines the interior. We then use a surface layer $\Sigma$ and the cut-and-paste approach to combine the two \cite{Lobo:2015lbc}.
 The metric in the interior region of the gravastar is a nonsingular de Sitter spacetime~\cite{Ovgun:2017jzt}:
\begin{equation}
ds^{2}=-\left(1-\frac{r_{-}^{2}}{\alpha^{2}}\right)dt_{-}^{2}+\left(1-\frac{r_{-}^{2}}{\alpha^{2}}\right)^{-1}dr_{-}^{2}+r_{-}^{2}d\Omega_{-}^{2},
\end{equation}
where $\alpha$ is the length of the AdS related to the cosmological constant $\Lambda$, $\alpha^2= - 3/\Lambda$. 

And the exterior of noncommutative geometry spacetime is given by~\cite{Campos:2021sff}:
\begin{equation}
ds^{2}=-f(r)_{+}dt_{+}^{2}+f(r)_{+}^{-1}dr_{+}^{2}+r_{+}^{2}d\Omega_{+}^{2}
\end{equation}
with 
\begin{equation} 
f(r)_{+}=1 - \frac{2 M}{r} + \frac{8M\sqrt{\theta}}{\sqrt{\pi}r^{2}}.
\end{equation} 
Note that $\pm$ stands for the exterior and interior geometry, respectively.

The resulting metrics are $g_{ij}^{+}$ and $g_{ij}^{-}$, in that order.  It is assumed that $g_{ij}^{+}(\xi)=g_{ij}^{-}(\xi)=g_{ij}(\xi)$, where $\xi^{i}=(\tau,\theta,\phi)$ is the hypersurface coordinates.  We want to create a single manifold ${M}$ by gluing ${ M_{+}}$ and ${ M_{-}}$ at their boundaries. This will make ${ M}={M_{+}}\cup{ M_{-}}$, and at the boundaries, $\Sigma=\Sigma_{+}=\Sigma_{-}$.

To calculate the stress-energy tensor components, we use the intrinsic metric on $\Sigma$ as follows:
\begin{equation} 
ds_{\Sigma}^{2}=-d\tau^{2}+a(\tau)^{2}\,(d\theta^{2}+\sin^{2}{\theta}\,d\phi^{2}).
\end{equation} 
Next, we use the Einstein field equation, $G_{{\mu}{\nu}}=8\pi\,T_{{\mu}{\nu}}$, where $c=G=1$.  Keep in mind that $x^{\mu}(\tau,\theta,\phi)=(t(\tau),a(\tau),\theta,\phi)$ is the junction surface.
One finds the unit normal vectors with respect to the junction surface are as follows \cite{Lobo:2015lbc}: 
\begin{equation} 
n_{-}^{\mu}=\left(\frac{1}{\left(1-\frac{a^{2}}{\alpha^{2}}\right)}\dot{a},\sqrt{\left(1-\frac{a^{2}}{\alpha^{2}}\right)+\dot{a}^{2}},0,0\right)\,,
\end{equation} 
\begin{eqnarray}
n_{+}^{\mu}=\left(\frac{1}{1 - \frac{2 M}{a} + \frac{8M\sqrt{\theta}}{\sqrt{\pi}a^{2}}}\dot{a},\sqrt{1 - \frac{2 M}{a} + \frac{8M\sqrt{\theta}}{\sqrt{\pi}a^{2}}+\dot{a}^{2}},0,0\right),
\label{normal}
\end{eqnarray}
where a derivative with regard to $\tau$ is represented by the overdot.
  The condition of the normal vectors for spherically symmetric spacetimes is $n^{\mu}n_{\mu}=+1$.  The following equation is used to determine the extrinsic curvatures \cite{MartinMoruno:2011rm}:
\begin{eqnarray}
K_{ij}^{\pm}=-n_{\mu}\left(\frac{\partial^{2}x^{\mu}}{\partial\xi^{i}\,\partial\xi^{j}}+\Gamma_{\;\;\alpha\beta}^{\mu\pm}\;\frac{\partial x^{\alpha}}{\partial\xi^{i}}\,\frac{\partial x^{\beta}}{\partial\xi^{j}}\right)\,.\label{extrinsiccurv}
\end{eqnarray}
so it is found as follows:
\begin{eqnarray}
K_{\;\;\psi}^{\psi\;-} & = & \frac{1}{a}\,\sqrt{\left(1-\frac{a^{2}}{\alpha^{2}}\right)+\dot{a}^{2}}\;,\label{genKplustheta-1}\\
K_{\;\;\tau}^{\tau\;-} & = & \left\{ \frac{\left(\ddot{a}-\frac{a}{\alpha^{2}}\right)}{\sqrt{\left(1-\frac{a^{2}}{\alpha^{2}}\right)+\dot{a}^{2}}}\right\} \,,\label{genKminustautau-1}
\end{eqnarray}
\begin{eqnarray}
K_{\;\;\psi}^{\psi\;+} & = & \frac{1}{a}\,\sqrt{1 - \frac{2 M}{a} + \frac{8M\sqrt{\theta}}{\sqrt{\pi}a^{2}}+\dot{a}^{2}}\;,\label{genKplustheta}
\end{eqnarray}
\begin{eqnarray}
K_{\;\;\tau}^{\tau\;+} & = & \left\{ \frac{ \ddot{a} + \dfrac{M}{a^2} + \dfrac{8M\sqrt{\theta}}{\sqrt{\pi}~a^3} }{ \sqrt{ 1 - \dfrac{2M}{a} + \dfrac{8M\sqrt{\theta}}{\sqrt{\pi}~a^2} + \dot{a}^2 } }\right\} \,,\label{genKminustautau}
\end{eqnarray}

The prime is for a derivative with respect to the $a$. Then we calculate the discontinuity as follows: $\kappa_{ij}=K_{ij}^{+}-K_{ij}^{-}$.

The stress-energy tensors $S_{\;j}^{i}$ on $\Sigma$ are calculated by following:
\begin{equation}
S_{\;j}^{i}=-\frac{1}{8\pi}\,\left(\kappa_{\;j}^{i}-\delta_{\;j}^{i}\;\kappa_{\;k}^{k}\right)\,.
\end{equation}
The surface energy density, $\sigma$, and the surface pressure, $\mathcal{P}$, may then be found using the relation $S_{\;j}^{i}={\rm diag}(-\sigma,\mathcal{P},\mathcal{P})$ as follows \cite{Lobo:2015lbc}:
\begin{widetext}
\begin{eqnarray}
\sigma & =-\frac{\kappa_{\;\psi}^{\psi}}{4\pi}= & -\frac{1}{4\pi a}\left[\sqrt{1 - \frac{2 M}{a} + \frac{8M\sqrt{\theta}}{\sqrt{\pi}~a^{2}}+\dot{a}^{2}}\;-\sqrt{\left(1-\frac{a^{2}}{\alpha^{2}}\right)+\dot{a}^{2}}\;\right],\label{gen-surfenergy2}\\
\mathcal{P} & =\frac{\kappa_{\;\tau}^{\tau}+\kappa_{\;\psi}^{\psi}}{8\pi}= & \frac{1}{8\pi a}\left[\frac{1+\dot{a}^{2}+a\ddot{a}-\frac{M}{a}}{\sqrt{1 - \frac{2 M}{a} + \frac{8M\sqrt{\theta}}{\sqrt{\pi}a^{2}}+\dot{a}^{2}}}-\frac{\left(1+a\ddot{a}+\dot{a}^2-\frac{2a^2}{\alpha^{2}}\right)}{\sqrt{\left(1-\frac{a^{2}}{\alpha^{2}}\right)+\dot{a}^{2}}}\right].\nonumber \\
\label{gen-surfpressure2}
\end{eqnarray}
Then it is found as follows:
\begin{eqnarray}
\sigma+2\mathcal{P} & =\frac{\kappa_{\;\tau}^{\tau}}{4\pi} & =\frac{1}{4\pi~a}\left[ \frac{ \ddot{a} + \dfrac{M}{a} - \dfrac{8M\sqrt{\theta}}{\sqrt{\pi}~a^2} }{ \sqrt{ 1 - \dfrac{2M}{a} + \dfrac{8M\sqrt{\theta}}{\sqrt{\pi}~a^2} + \dot{a}^2 }} - \frac{\left(\ddot{a}-\frac{a^2}{\alpha^{2}}\right)}{\sqrt{\left(1-\frac{a^{2}}{\alpha^{2}}\right)+\dot{a}^{2}}} \right].\nonumber \\
\label{s2P}
\end{eqnarray}
\end{widetext}

To calculate the surface mass of the thin-shell, one can use this equation: $M_{s}(a)=4\pi a^{2}\sigma$. To find a stable solution, we consider a static case [$a_{0}\in(r_{-},r_{+})$].
\begin{widetext}
Then the surface energy density and pressure at static case reduce to  
\begin{eqnarray}
\sigma(a_{0}) & = & -\frac{1}{4\pi a_{0}}\left[\sqrt{1-\frac{2M}{a_{0}}+\frac{8M\sqrt{\theta}}{\sqrt{\pi}~a_{0}^{2}}}
-\sqrt{1+\frac{\Lambda a_{0}^{2}}{3}}\right],\label{gen-surfenergy2a}
\end{eqnarray}
\begin{eqnarray}
\mathcal{P}(a_{0}) & = & \frac{1}{8\pi a_{0}}\left[\frac{1-\frac{M}{a_{0}}}{\sqrt{1-\frac{2M}{a_{0}}+\frac{8M\sqrt{\theta}}{\sqrt{\pi}~a_{0}^2}}}-\frac{\left(1+\frac{2\Lambda a_{0}^2}{3}\right)}{\sqrt{1+\frac{\Lambda a_{0}^{2}}{3} }}\right]. 
\label{gen-surfpressure2a}
\end{eqnarray}
Then one can write that 
\begin{eqnarray}
\sigma(a_{0})+2\P(a_{0}) & = & \frac{1}{4\pi a_0}\left[ \frac{\dfrac{M}{a_{0}} - \dfrac{8M\sqrt{\theta}}{\sqrt{\pi}~a_{0}^2} }{ \sqrt{ 1 - \dfrac{2M}{a_{0}} + \dfrac{8M\sqrt{\theta}}{\sqrt{\pi}~a_{0}^2}}} - \frac{\frac{\Lambda a_{0}^2}{3 }}{\sqrt{1+\frac{\Lambda a_{0}^{2}}{3} }} \right].
\end{eqnarray}
The dimensionless quantities, $\Tilde{\Lambda} = {\Lambda a_0^2}/{3}$, $\Theta = {8\sqrt{\theta}}/{(\sqrt{\pi}M)}$ and $\Tilde{a} = {a_{0}}/{ M}$ are now introduced, allowing the aforementioned equations to adopt the following form:

\begin{eqnarray}
\Tilde{\sigma} & = & -\frac{1}{4\pi}\left[\sqrt{1- \dfrac{2}{\Tilde{a}} + \dfrac{\Theta}{~\Tilde{a}^2}}-\sqrt{1 +\Tilde{\Lambda}}\right],\label{gen-surfenergyadm}
\end{eqnarray}

\begin{figure}[h!]
\includegraphics[scale=0.35]{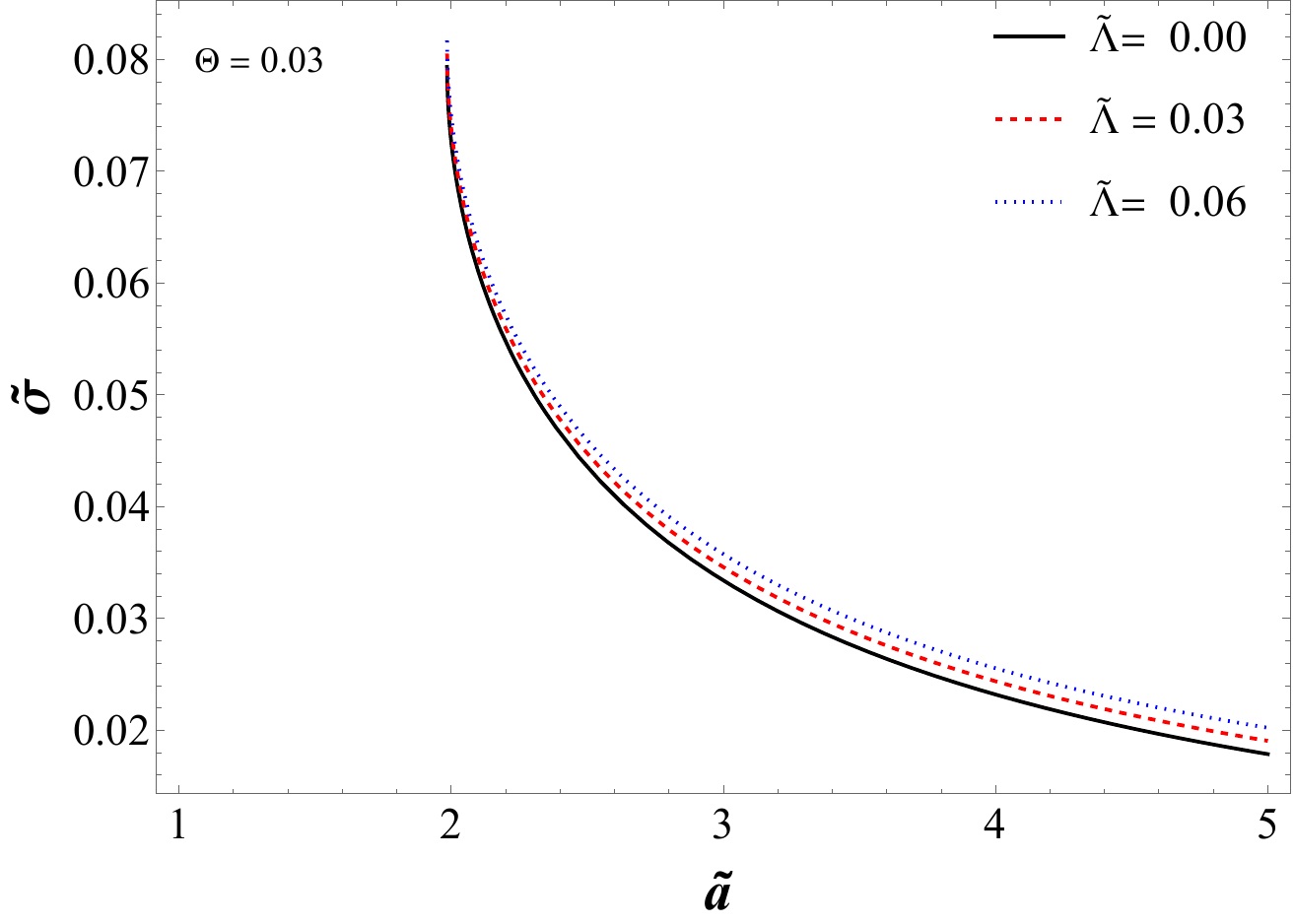} 
\caption{Dimensionless surface energy density~$\Tilde{\sigma}$ as a function of the dimensionless shell radius $\Tilde{a}$ for fixed noncommutative parameter $\Theta = 0.03$ and different values of the effective cosmological parameter, $\Tilde{\Lambda} = 0$ (black, solid), $\Tilde{\Lambda} = 0.03$ (red, dashed) and $\Tilde{\Lambda} = 0.06$ (blue, dotted).}
\label{EnergyDenAdm}
\end{figure}
{In Figure (\ref{EnergyDenAdm}), the chosen value of $\Theta$ represents a regime where noncommutative corrections are significant while remains within the perturbative approximation adopted in this study. Increasing $\Tilde{\Lambda}$ raises the surface energy density across the entire range of shell radius, illustrating the combined contribution of the cosmological constant and noncommutative effects on the shell structure.}
\begin{eqnarray}
\Tilde{\mathcal{P}} & = & \frac{1}{8\pi}\left[\frac{1-\frac{1}{\Tilde{a}}}{\sqrt{1-\frac{2}{\Tilde{a}}+\frac{\Theta}{~\Tilde{a}^{2}}}}-\frac{\left(1 + 2 \Tilde{\Lambda}\right)}{\sqrt{1 + \Tilde{\Lambda}}}\right]. 
\label{gen-surfpressureadm}
\end{eqnarray}

\begin{figure}[h!]
\includegraphics[scale=0.35]{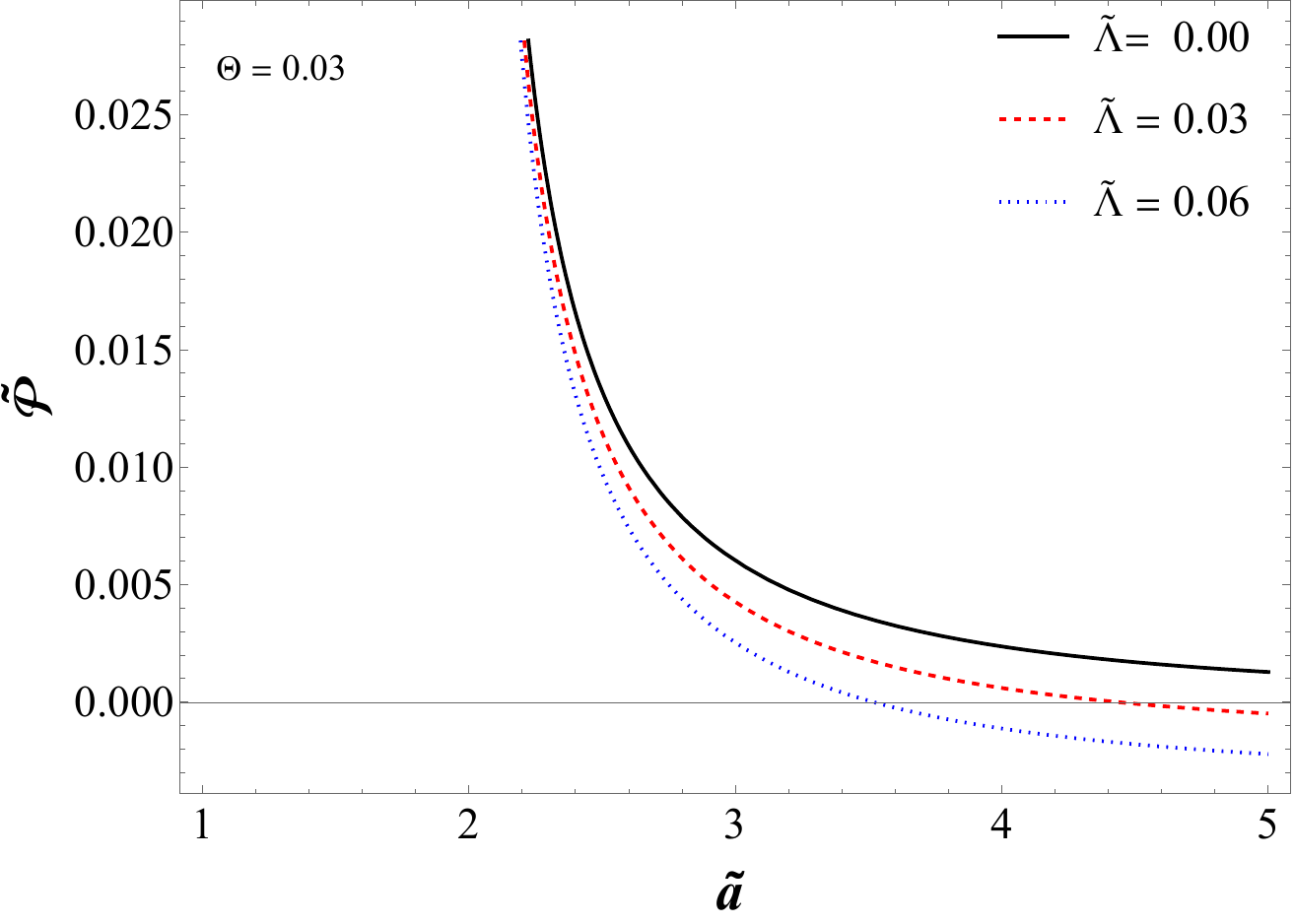} %
\caption{Dimensionless surface pressure $\tilde{\mathcal{P}}$ as a function of the shell radius $\Tilde{a}$ for $\Theta = 0.03$ and $\Tilde{\Lambda} = 0$ (black, solid), $\Tilde{\Lambda} = 0.03$ (red, dashed) and $\Tilde{\Lambda} = 0.06$ (blue, dotted).}
\label{PressureADM}
\end{figure}

 \end{widetext}
 
{In Figure (\ref{PressureADM}), the selected values of $\Tilde{\Lambda}$ allow one to isolate the influence of the effective cosmological constant while keeping the noncommutative contribution fixed. Larger values of $\Tilde{\Lambda}$ increase the pressure required to maintain equilibrium of the thin shell.}
\begin{figure}[h!]
\includegraphics[scale=0.33]{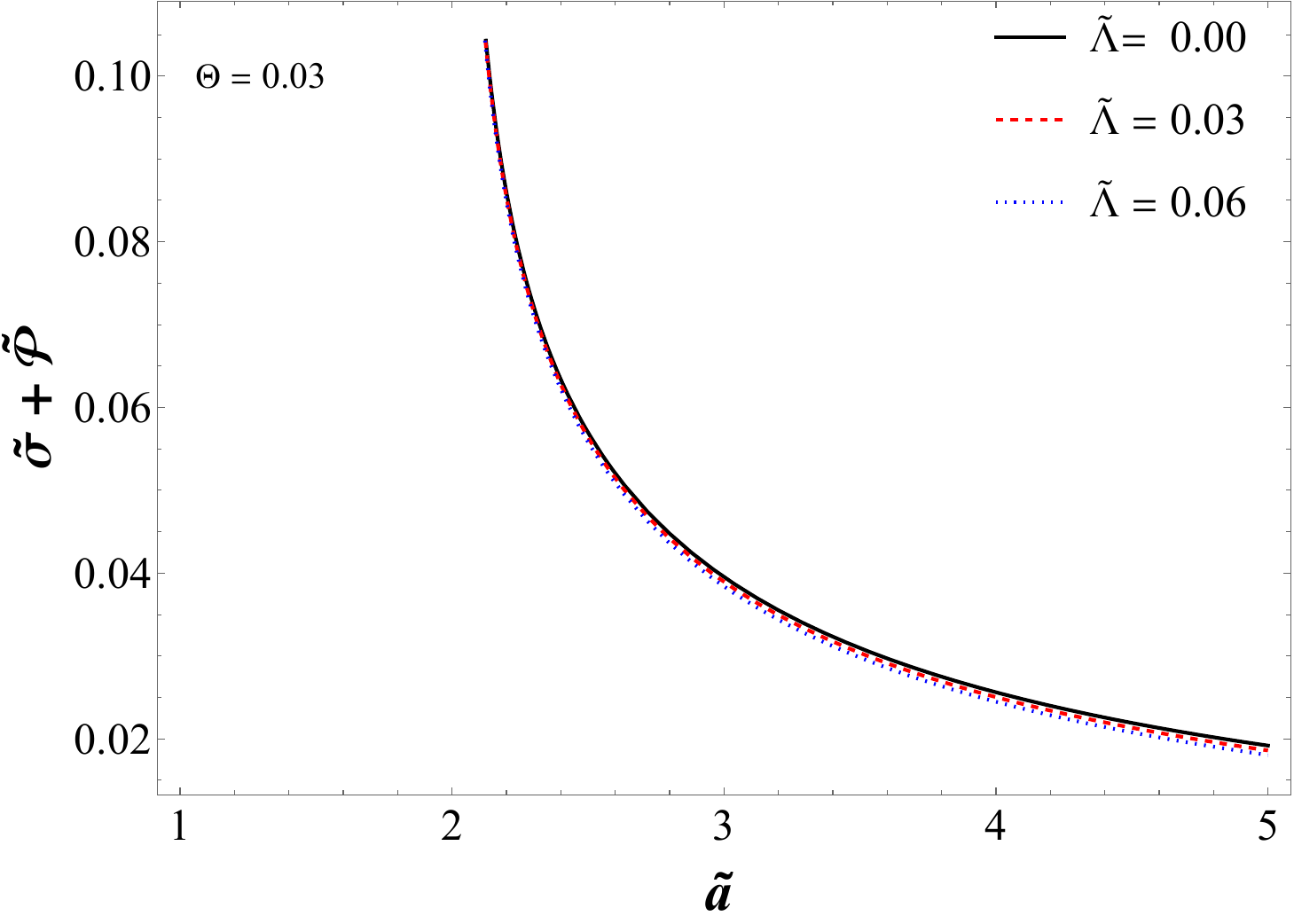}
\caption{Profile of the dimensionless combination $\Tilde{\sigma} + \Tilde{\mathcal{P}}$, which determines the weak and null energy conditions, as a function of the shell radius $\Tilde{a}$. The calculations are performed for $\Theta = 0.03$ and  $\Tilde{\Lambda} = 0$ (black, solid), $\Tilde{\Lambda} = 0.03$ (red, dashed) and $\Tilde{\Lambda} = 0.06$ (blue, dotted). The positivity of the curves throughout the displayed interval confirms that both the weak and null energy conditions are satisfied within the adopted parameter range.}
\label{3dGrad}
\end{figure}

Then one can write that $\left[\sigma(a_{0})+2\mathcal{P}(a_{0})\right] a_0 = \Tilde{\sigma}+2\Tilde{\mathcal{P}}$ 
\begin{eqnarray}
\Tilde{\sigma}+2\Tilde{\mathcal{P}} & = & \frac{1}{4\pi}\left[ \frac{\dfrac{1}{\Tilde{a}} - \dfrac{\Theta}{~\Tilde{a}^2} }{ \sqrt{1-\frac{2}{\Tilde{a}}+\frac{\Theta}{~\Tilde{a}^{2}}}} - \frac{\Tilde{\Lambda}}{\sqrt{1+\Tilde{\Lambda}}} \right].
\end{eqnarray}

\begin{figure}[h!]
\includegraphics[scale=0.37]{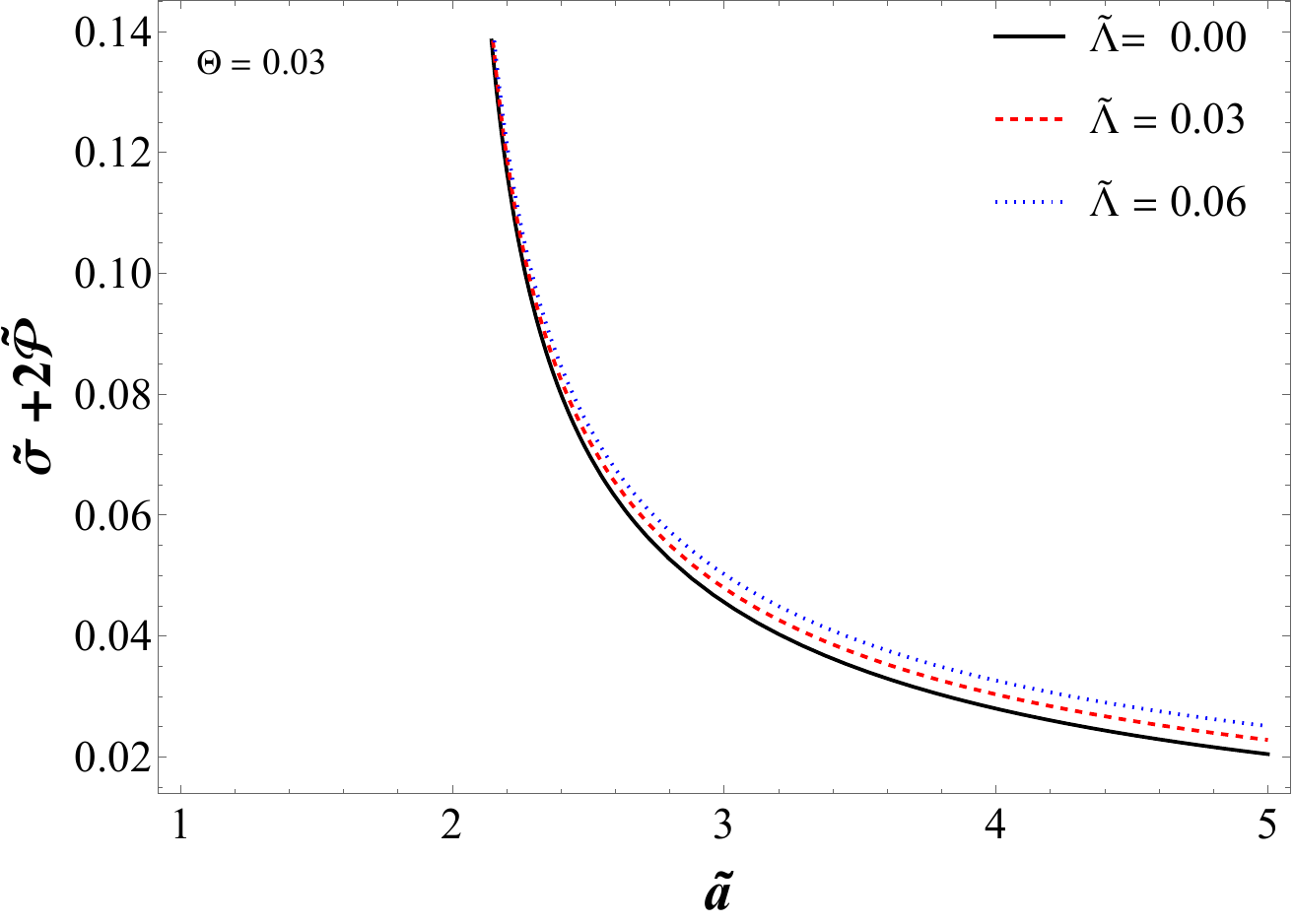} 
\caption{Dimensionless quantity $\Tilde{\sigma} + 2\Tilde{\mathcal{P}}$ as a function of $\Tilde{a}$ for fixed $\Theta = 0.03$ and  $\Tilde{\Lambda} = 0$ (black, solid), $\Tilde{\Lambda} = 0.03$ (red, dashed) and $\Tilde{\Lambda} = 0.06$ (blue, dotted). This combination is relevant for the strong energy condition. The positive values found for all curves indicate that the strong energy condition remains fulfilled within the considered parameter domain.}
\label{Enrgy+2P}
\end{figure}

The energy conditions state that if $\sigma \geq 0$ and $\sigma + \mathscr{P} \geq 0 $ are satisfied, then the weak energy condition (WEC) must exist.  Since $\sigma + \mathcal{P}\geq 0 $, the null energy condition (NEC) is valid by continuity.  $\sigma + \mathscr{P} \geq 0$ and $\sigma + 2 \mathscr{P} \geq 0$~\cite{Horvat:2007qa} must both be true to validate the strong energy condition (SEC).  The pressure $\mathscr{P}$ (Fig. \ref{PressureADM}) and $\sigma$ (Fig. \ref{EnergyDenAdm}) are both positive, as our calculations show.  The positive pressure and energy density maintain $\tilde{\sigma}+\tilde{\mathcal{P}} \geq 0$ (Fig. \ref{3dGrad}) and $\tilde{\sigma}+ 2\tilde{\mathcal{P}} \geq 0$ (Fig. \ref{Enrgy+2P}) for sufficiently small $\Theta$.  Therefore, the energy criteria are satisfied when $\Theta$ is small. Additionally, we observe that these requirements are still satisfied even when we consider $\Tilde{\Lambda} = 0$ because the noncommutativity parameter $\Theta$ controls the geometry and maintains divergences that are typical of classical solutions from appearing. This suggests that even in the absence of an effective cosmological constant, the effects of noncommutative gravity are crucial for maintaining the stability of the gravastar shell.

Now, for $\Theta\ll 1\, (\theta\ll 1)$ and $a\gg 1$, with $\Lambda=0$ and $\theta\neq 0$, we find

\begin{eqnarray}
\tilde{\sigma}_{\theta}=\frac{1}{4\pi}\left(\dfrac{1}{\tilde{a}} + \frac{1}{2\tilde{a}^{2}} - \frac{\Theta}{2\tilde{a}^2} \right), 
\label{sth}
\end{eqnarray}
\begin{eqnarray}
\tilde{\mathcal{P}}_{\theta}=\frac{1}{8\pi}\left(\frac{1}{2\tilde{a}^{2}} - \frac{\Theta}{2\tilde{a}^2}\right), 
\label{pth}
\end{eqnarray}

for the limit $\Lambda \gg1$, $\theta= 0$, we have

\begin{eqnarray}
\tilde{\sigma}_{\Lambda}=\frac{1}{4\pi}\left(\dfrac{1}{\tilde{a}}  + \sqrt{\tilde{\Lambda}}+\dfrac{1}{2\tilde{a}^{2}} \right), 
\label{sl1}
\end{eqnarray}
\begin{eqnarray}
\tilde{\mathcal{P}}_{\Lambda}=\frac{1}{8\pi}\left(\frac{1}{2\tilde{a}^2} -2\sqrt{\tilde{\Lambda}}\right), 
\label{pl1}
\end{eqnarray}

\begin{figure}[htbh!]
\centering
\subfigure[]{\includegraphics[scale=0.17]{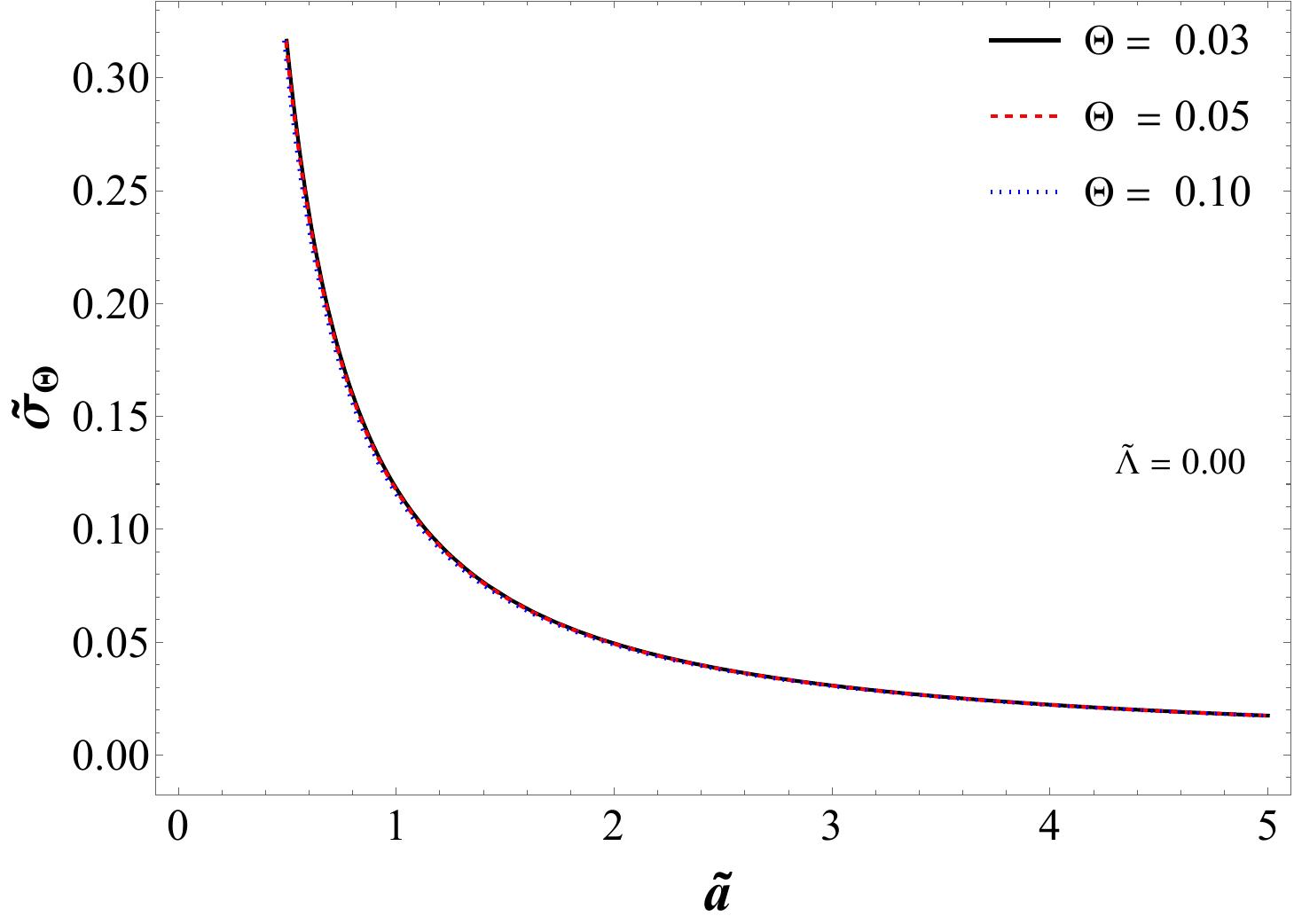}}
\subfigure[]{\includegraphics[scale=0.17]{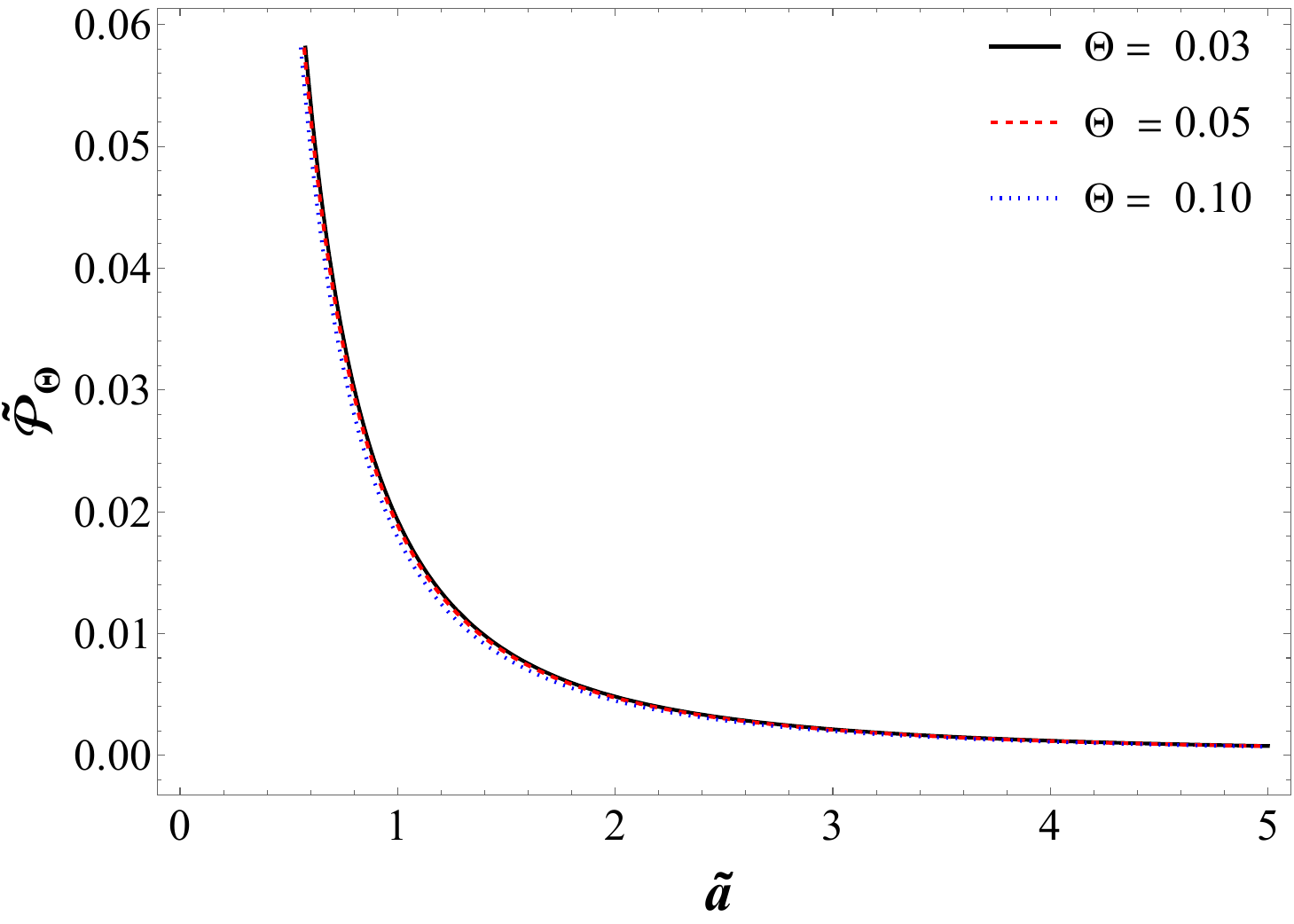}}
\subfigure[]{\includegraphics[scale=0.17]{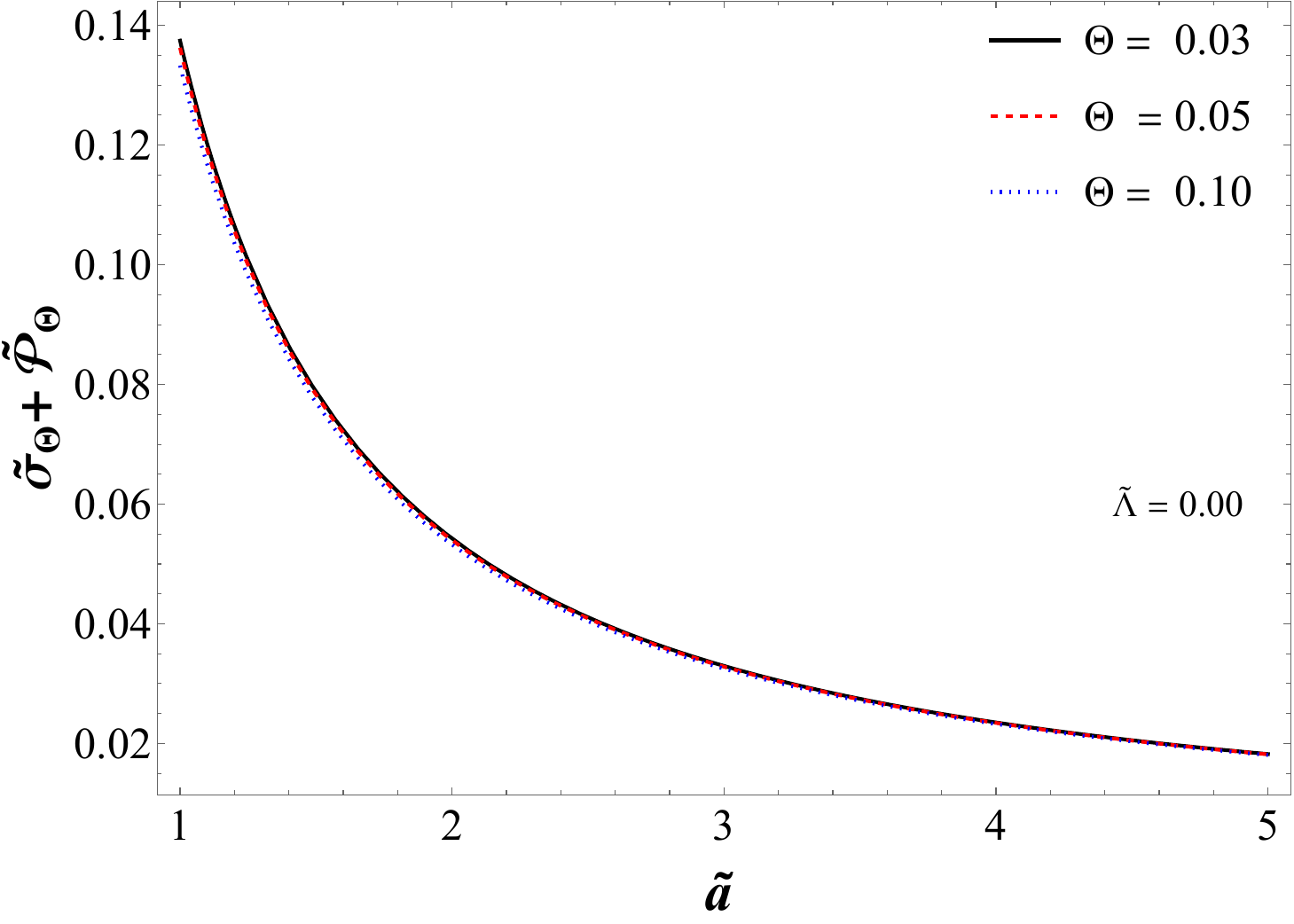}}
\subfigure[]{\includegraphics[scale=0.17]{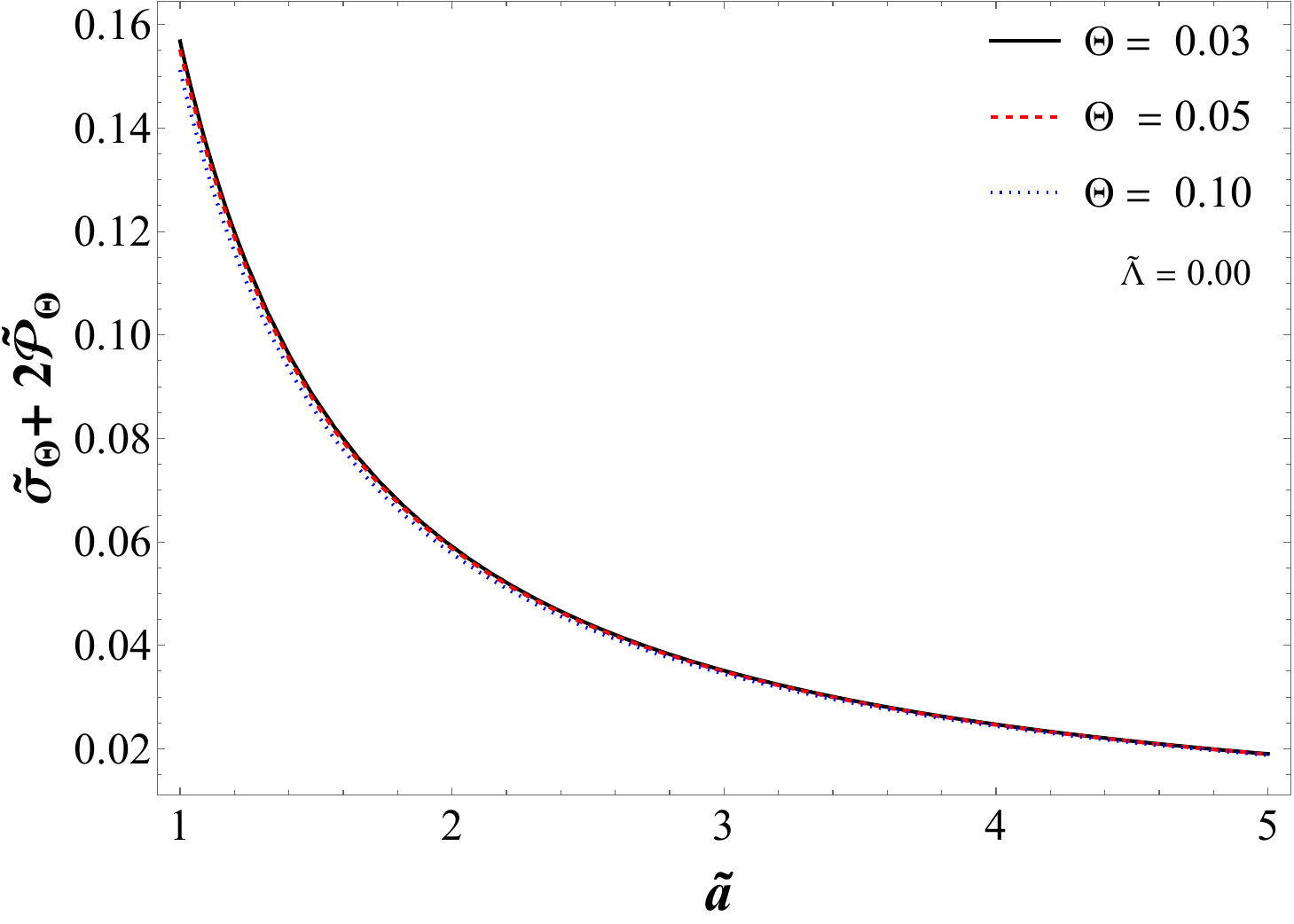}}
\caption{Energy condition functions for different values of the noncommutative parameter, $\Theta= 0.03$, $0.05$, and $0.10$, with the effective cosmological parameter fixed at $\Tilde{\Lambda}=0$. Panels show (a) the surface energy density $\Tilde{\sigma}_{\Theta}$, (b) the surface pressure $\tilde{\mathcal{P}}_{\Theta}$, (c) $\Tilde{\sigma}_{\Theta}+\tilde{\mathcal{P}}_{\Theta}$ and (d) $\Tilde{\sigma}_{\Theta}+2\tilde{\mathcal{P}}_{\Theta}$. The values chosen for $\Theta$ cover the physically relevant range of noncommutative effects. Our results show that noncommutativity is sufficient to satisfy the energy conditions, highlighting its role as an effective source analogous to the cosmological constant.}
\label{EnergyConditions2}
\end{figure}
\begin{figure}[htbh!]
\centering
\subfigure[]{\includegraphics[scale=0.17]{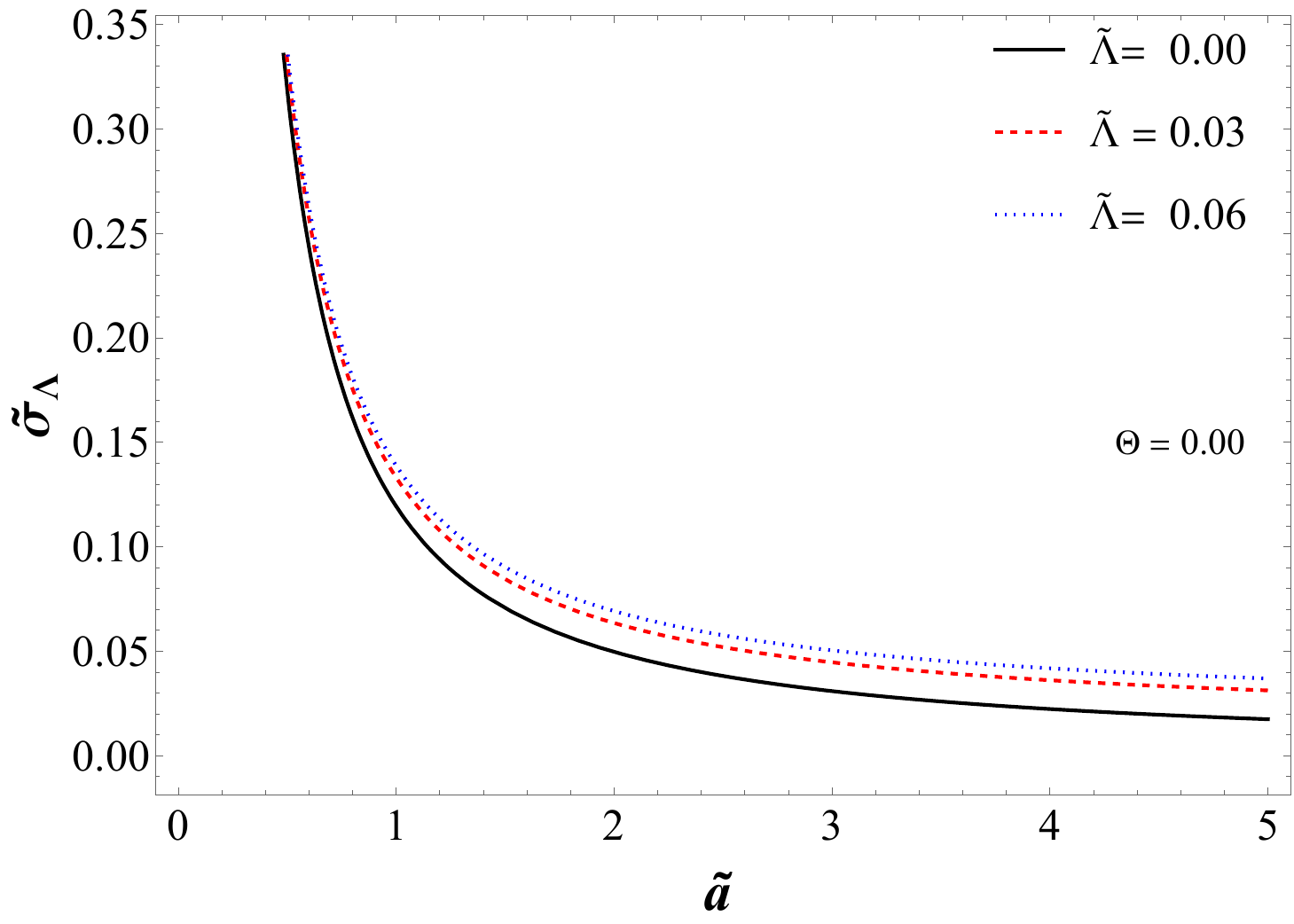}}
\subfigure[]{\includegraphics[scale=0.17]{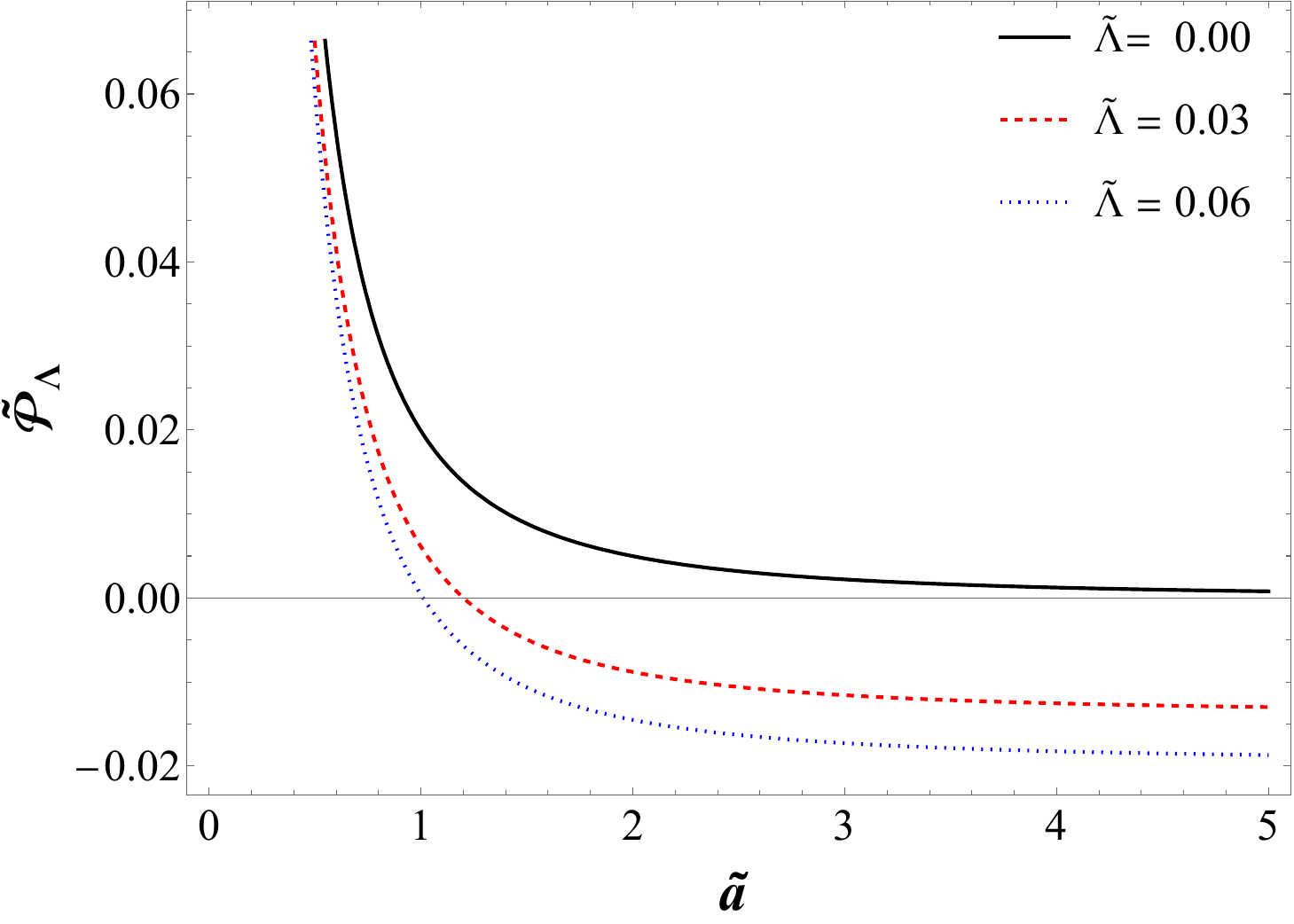}}
\subfigure[]{\includegraphics[scale=0.17]{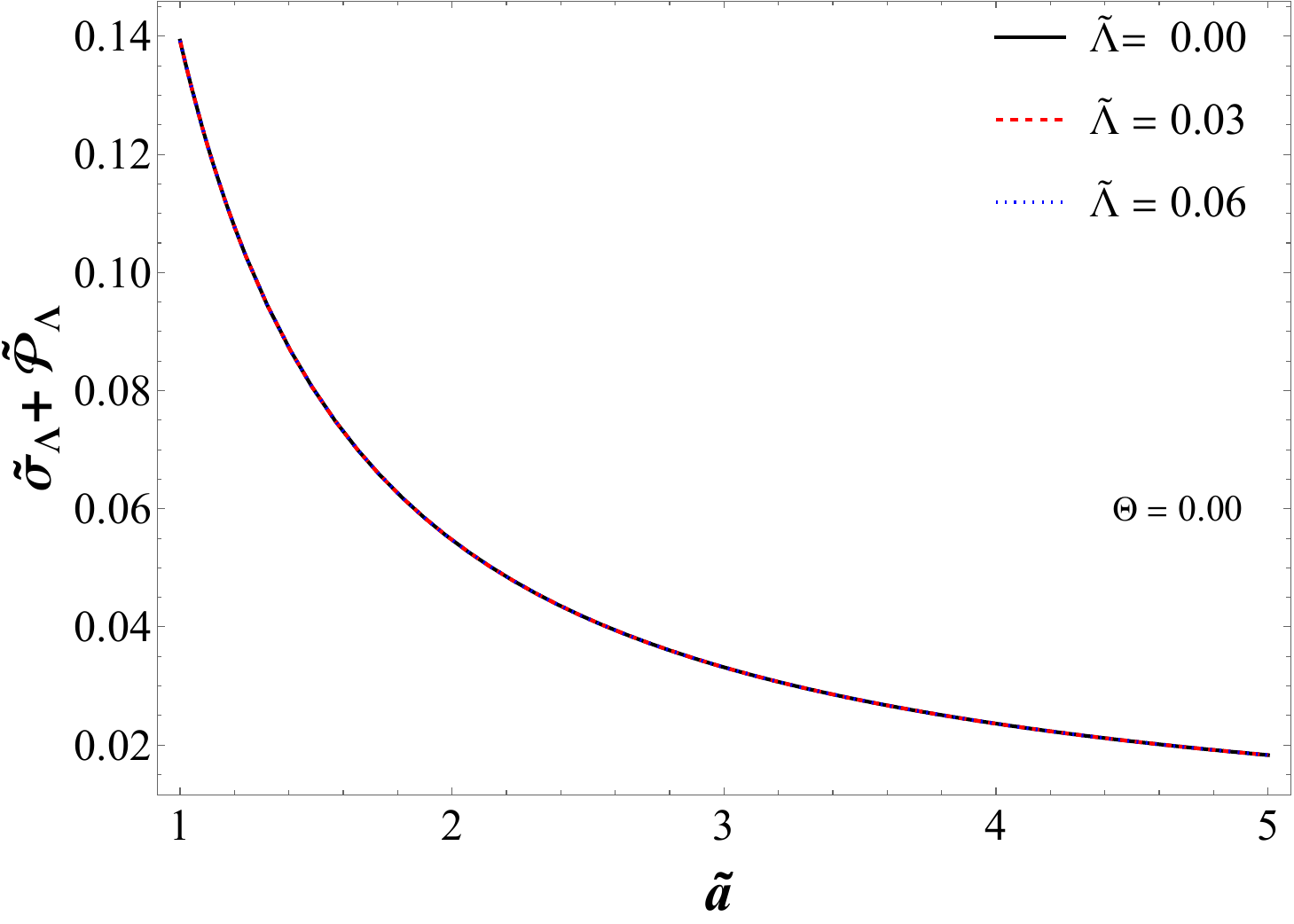}}
\subfigure[]{\includegraphics[scale=0.17]{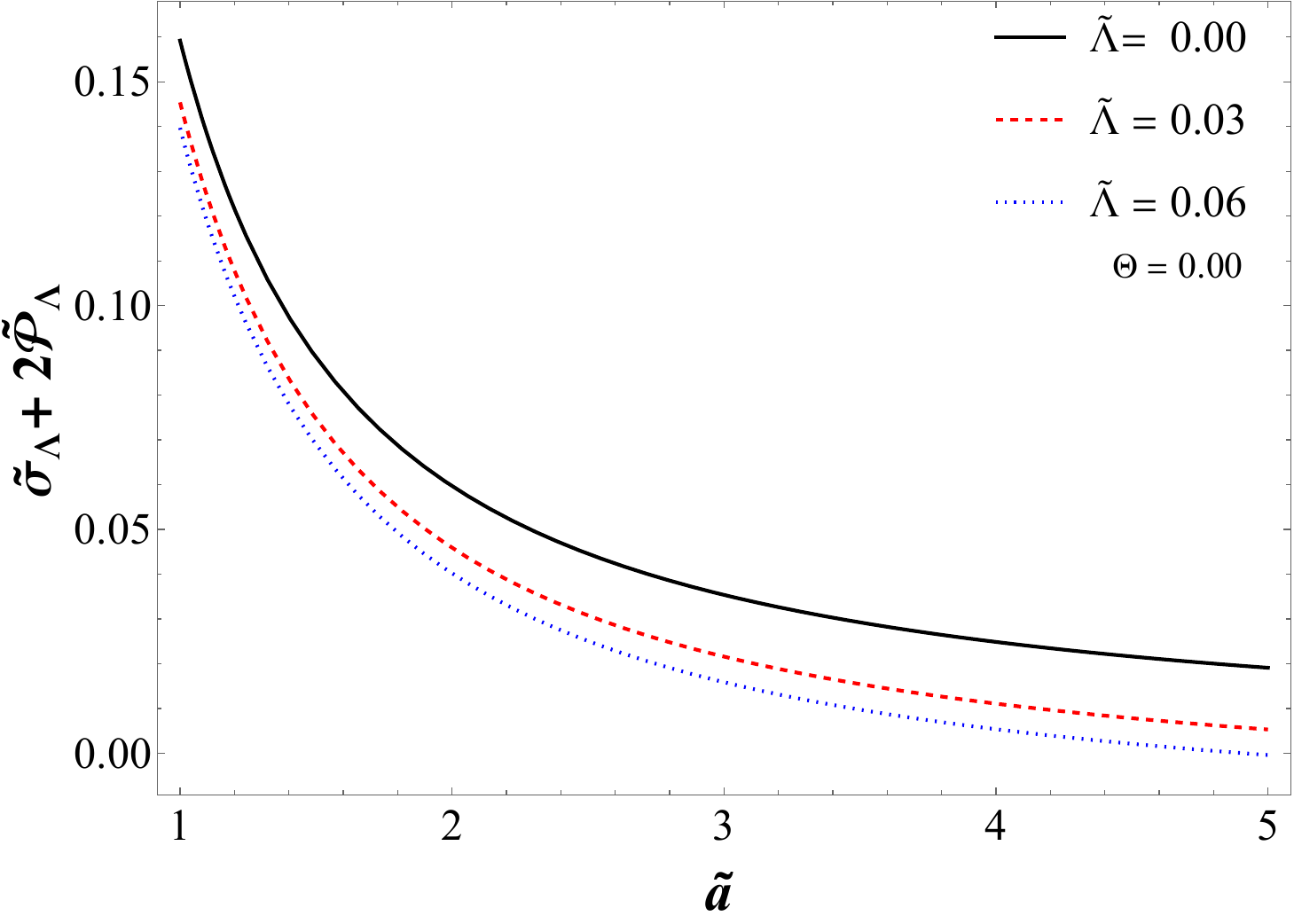}}
\caption{Energy condition functions obtained in the noncommutative are null $\Theta = 0$. Panels show (a) the surface energy density $\Tilde{\sigma}_{\Lambda}$, (b) the surface pressure $\tilde{\mathcal{P}}_{\Lambda}$, (c) $\Tilde{\sigma}_{\Lambda}+\tilde{\mathcal{P}}_{\Lambda}$ and (d) $\Tilde{\sigma}_{\Lambda}+2\tilde{\mathcal{P}}_{\Lambda}$, for different values of $\tilde{\Lambda}$. This figure provides a reference configuration in which only the cosmological contribution is present, allowing for a direct comparison with the noncommutative case presented earlier.}
\label{EnergyConditions}
\end{figure}
{To better understand the noncommutativity parameter $\Theta$ related to the gravastar structure and its possible connection with the cosmological constant, we begin with an analysis of the energy conditions, considering $\Lambda = 0$. In this context, the equations for the surface energy density and pressure are given by Eqs. (\ref{sth}) and (\ref{pth}), respectively.}

{Additionally, we analyze the case where the noncommutative effects vanish $\Theta=0$, retaining only the contribution of the cosmological constant. Equations (\ref{sl1}) and (\ref{pl1}) show the corresponding expressions for energy density and pressure. The results presented in Fig. (\ref{EnergyConditions}) indicate behavior qualitatively similar to that observed in the noncommutative case, suggesting that both $\Theta$ and $\Lambda$ may have similar effects on energy conditions and the stability of the configuration.}

{Nevertheless, it meets the SEC, since $\tilde{\sigma} + 2\tilde{\mathcal{P}} > 0$.}
Note that for $\tilde{\mathcal{P}_{\Lambda}}>0$, we have the following condition $\tilde{a}>1/2(\tilde{\Lambda})^{1/4}$.

Therefore, a relationship between $\theta$ and $\Lambda$ can be obtained, such that
\begin{eqnarray}
\Theta=4\tilde{a}^2\sqrt{\tilde{\Lambda}}, \quad \mbox{or}  \quad 
{\theta}=\frac{12\pi\Lambda a^4_0}{\mu^2}.
\label{tetalambda1}
\end{eqnarray}

Now admitting $\mu =12M/a_0= {M_{BH}}/{M_{\odot}}$, where $M_{BH} $ is the mass of the black hole and 
$M_{\odot}=1.989 \times 10^{30}$ kg is the solar mass,
then, for $M_{BH} =10M_{\odot}$, $a_0\approx 29.5 \times 10^{3}$ m (radius of the black hole) and cosmological constant $\Lambda=1.088 \times 10 ^{-58}$ m$^{-2}$, we obtain the following value for the parameter $\theta$:
\begin{eqnarray}
\theta \sim 
\left[10\, TeV \right]^{-2}
\qquad \mbox{or} \qquad \sqrt{\theta}\sim\left[10\, TeV \right]^{-1}.
\end{eqnarray}
Thus, we obtain the following energy scale: $\Lambda_{NC}\sim 1/\sqrt{\theta}\sim 10\, TeV $. The results obtained are in full agreement with the existing literature~\cite{Mocioiu:2000ip,Falomir:2002ih,Vagnozzi:2022moj} and, in particular, are compatible with the results derived within the framework of the gravastar model in noncommutative and  minimal length BTZ geometry~\cite{silva2024,Anacleto:2025oef}.

{For $\tilde{\Lambda}=0$, the energy conditions remain unchanged, as shown in Fig.~(\ref{EnergyConditions2}), indicating that noncommutativity effectively reproduces the contribution otherwise associated with the cosmological constant.}

\section{Null Geodesic Equations}\label{sec3}

The null geodesic equations for the static spacetime of the previously acquired  will be derived in this part, together with the boundary conditions.  The null vector is represented by $k^\mu=dx^\mu/d\lambda$, and the affine parameter by $\lambda$.  Consequently, the geodesic equations are frequently shown as \cite{Sakai:2014pga}
\begin{eqnarray}
{dk^\mu\over d\lambda}+\Gamma^\mu_{\nu\rho}k^\nu k^\rho=0,~~~
{\rm with}~~~k_\mu k^\mu=0.
\label{geodesic}
\end{eqnarray}
The geodesics in the $\theta=\pi/2$ plane for the outside $(+)$ and the inside $(-)$ are given by
\beq\label{geo1}{
{d\over d\lambda_\pm}(f(r)_\pm k^t_\pm)=0,~~~
{d\over d\lambda_\pm}(r_\pm^2k^\varphi_\pm)=0,}
\eeq\beq
{1\over\sqrt{f(r)_\pm}}{d\over d\lambda_\pm}\left({k_\pm^r\over\sqrt{f(r)_\pm}}\right)+{df(r)_\pm\over dr_\pm}{(k^t_\pm)^2\over2}-r_\pm(k^\varphi_\pm)^2=0,
\label{geo2}\eeq
\beq
-f(r)_\pm(k^t_\pm)^2+{(k^r_\pm)^2\over f(r)_\pm}+r_\pm^2(k^\varphi_\pm)^2=0.
\label{geo3}\eeq
Because Eq.(\ref{geo2}) is also derived from (\ref{geo1}) and (\ref{geo3}), we do not have to solve it.
Equations (\ref{geo1}) are integrated as
\beq\label{geo11}
f(r)_\pm k^t_\pm={\rm const}.\equiv E_\pm,~~~
r_\pm^2k^\varphi_\pm={\rm const}.\equiv L_\pm,
\eeq
and then (\ref{geo3}) becomes
\beq\label{geo21}
(k_\pm^r)^2+{f(r)_\pm L_\pm^2\over r_\pm^2}=E_\pm^2.
\label{geo31}\eeq
It follows from (\ref{geo11}) and (\ref{geo21}) that
\beq
{dr_\pm\over d\varphi}={k^r_\pm\over k^\varphi_\pm}={r_\pm^2k_\pm^r\over L_\pm}
=\pm r_\pm\sqrt{\left({E_\pm r_\pm\over L_\pm}\right)^2-f(r)_\pm},
\label{geo42}\eeq
which gives null geodesics in the exterior and interior regions of the gravastar.

The equation for the interior region in (\ref{geo42}) is integrated as
\beq
r_-=r_m\sec(\varphi-\varphi_m),~~~
r_m\equiv\left({E_-^2\over L_-^2}+ \tilde{\Lambda}^2\right)^{-\frac12},
\eeq
where an integral constant is $\varphi_m$.
 Since the gravastar's surface is transparent in our scenario and there is no black hole horizon, any incident light beam must pass through it.   Thus, there are two points where the penetrating null geodesic intersects the gravastar $\Sigma$ surface.
 The $\varphi$-coordinate values of those two crossing sites, $\varphi_1$ and $\varphi_2~(\varphi_1 < \varphi_2$), are ascertained via
\beq
\varphi_m=\varphi_1+\arccos{r_m\over a_0}=\varphi_2-\arccos{r_m\over a_0}.
\eeq
However, it is not possible to integrate analytically the equation for the exterior region in (\ref{geo42}).
 However, by entering $f(r)_+\rightarrow1$, the asymptotic solution at $r\rightarrow\infty$ is obtained:
\beq\label{asysol}
{r_+={L_+\over E_+}}\sec(\varphi-\varphi_c),
\eeq
where $\varphi_c$ is an integral constant.

Since, in our case, the gravastar is static, we proceed to analyze the boundary conditions $k^\mu$ in $\Sigma$. Thus, we obtain
\beq\label{kbc1}
\sqrt{f(a_0)_+}k^t_+=\sqrt{f(a_0)_-}k^t_-,~~~k^\varphi_+=k^\varphi_-.
\eeq
With the help of the null condition (\ref{geo3}), we also obtain
\beq\label{kbc2}
{k^r_+\over\sqrt{f(a_0)_+}}={k^r_-\over\sqrt{f(a_0)_-}}.
\eeq
The continuity condition along the shell, combined with the integration constants specified in (\ref{geo11}) and (\ref{kbc1}),
\beq\label{kbc3}
L_+=L_-,~~~{E_+\over\sqrt{f(a_0)_+}}={E_-\over\sqrt{f(a_0)_-}}.
\eeq
To simplify notation, we will adopt $L$ to represent both $L_+$ and $L_-$, since both are identical.

To make a qualitative discussion on photon trajectories, it is convenient to introduce the effective potential as follows.
Equation (\ref{geo21}) is rewritten as {
\beq\label{geo33}
\left({dr_\pm\over d\lambda_\pm}\right)^2+{L^2f(r)_\pm\over r_\pm^2}=E_\pm^2.
\eeq
To discuss the dynamics with a continuous ``potential" by analogy with the Newtonian mechanics, we introduce unified variables a
\begin{equation}
\begin{aligned}
& r = r_- , \quad \lambda = \lambda_- = \sqrt{\frac{f(a_0)_-}{f(a_0)_+} }\quad \text{(inside)},\\
& r = r_+ , \quad \lambda = \lambda_+ \quad \text{(outside)},
\end{aligned}
\end{equation}
and we define the effective potential as
\begin{equation}
V(r) =
\begin{cases}
 \dfrac{L^2 f(r)_-}{r^2} \dfrac{f(a_0)_+}{f(a_0)_-} = 
 \dfrac{L^2f(a_0)_+}{r^2f(a_0)_-} \left( 1-\dfrac{(r-a_0)^{2}}{\alpha^{2}} \right), & r < a_0 \\[4mm]
\dfrac{L^2 f(r)_+}{r^2} = \dfrac{L^2}{r^2} \left( 1 - \dfrac{2M}{a_0 +r} + \dfrac{8M \sqrt{\theta}}{\sqrt{\pi}~(a_0 +r)^{2}} \right), & r > a_0~.
\end{cases}
\end{equation}
Now, we can rewrite 
\begin{equation}
V(r)(r^2/L) =
\begin{cases}
 \dfrac{Lf(a_0)_+}{f(a_0)_-} \left( 1 +\Tilde{\Lambda}\right), & r < a_0 \\[4mm]
L \left( 1- \dfrac{1}{\Tilde{a}_0} + \dfrac{\Theta}{~\Tilde{a}_0} \right), & r > a_0~.
\end{cases}
\label{eff}
\end{equation}
Then we obtain the continuous equation of motion,
\beq
\left({dr\over d\lambda}\right)^2+V(r)=E_+^2.
\eeq

\begin{figure}[h!]
\includegraphics[scale=0.4]{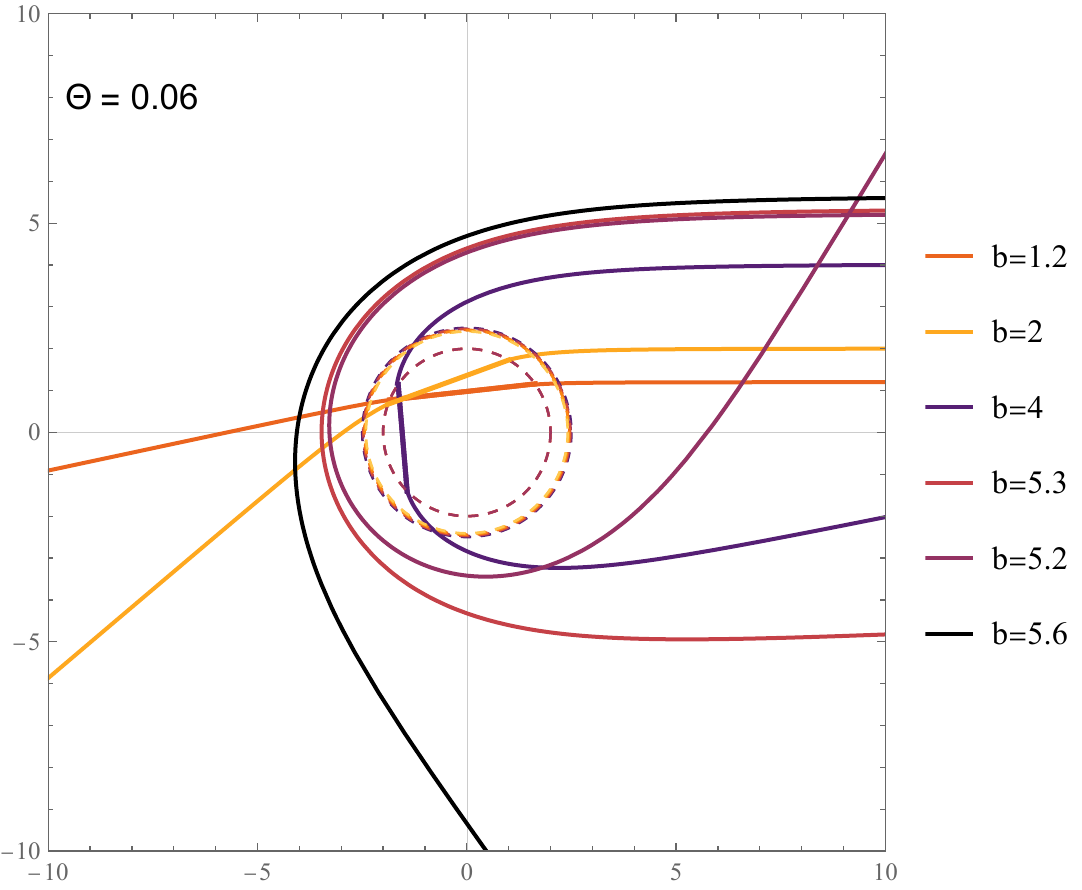} 
\caption{Photon trajectories around a transparent noncommutative thin-shell gravastar for $\Theta=0.06$. The different curves correspond to the impact parameters $b = 1.2 , 2, 4, 5.2, 5.3$ and $5.6$, selected to illustrate the transition from strongly penetrating trajectories to weakly scattered ones, including trajectories close to the critical impact parameter $b_c$. Unlike the black-hole case, photons with $b<b_
c$ are not absorbed but traverse the transparent shell and re-emerge, producing observational signatures characteristic of gravastars.}
 \label{Veff}
 \end{figure}

\begin{figure}[htbh!]
\centering
\subfigure[]{\includegraphics[scale=0.25]{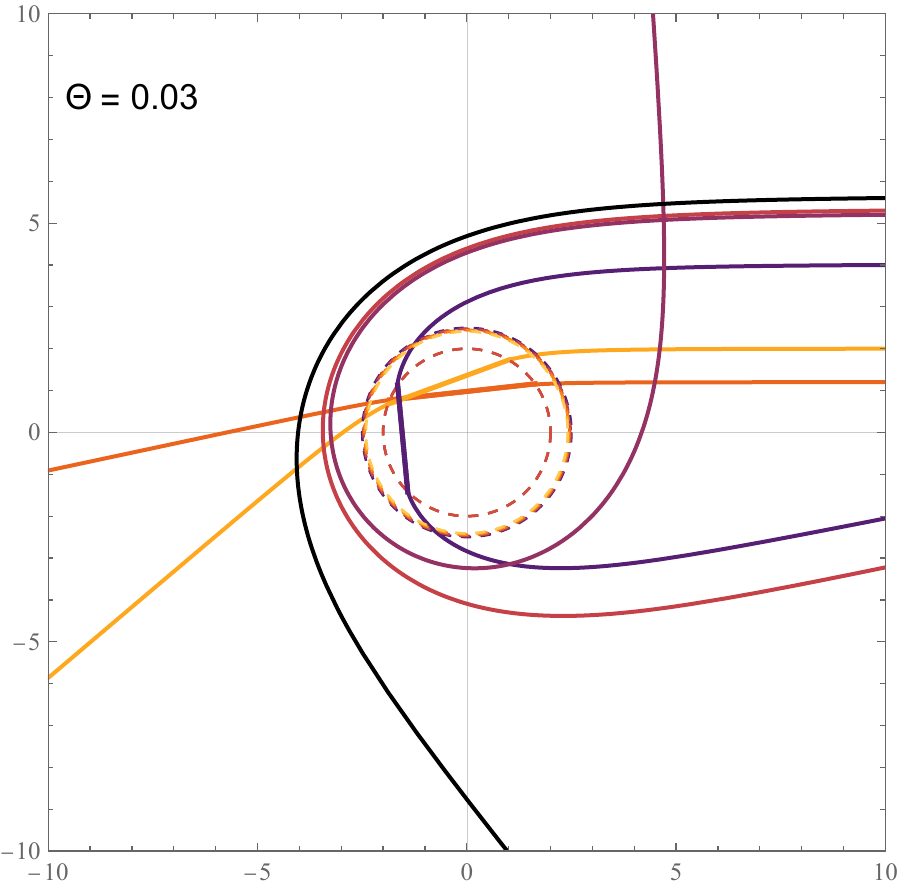}}
\subfigure[]{\includegraphics[scale=0.25]{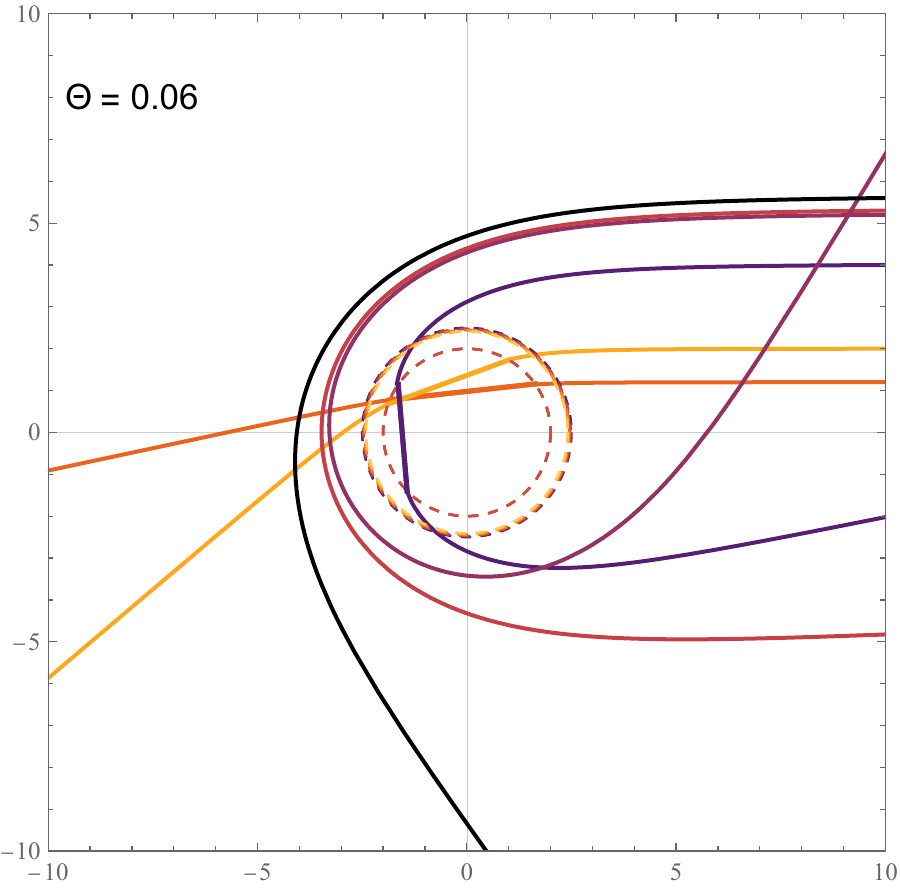}}
\subfigure[]{\includegraphics[scale=0.25]{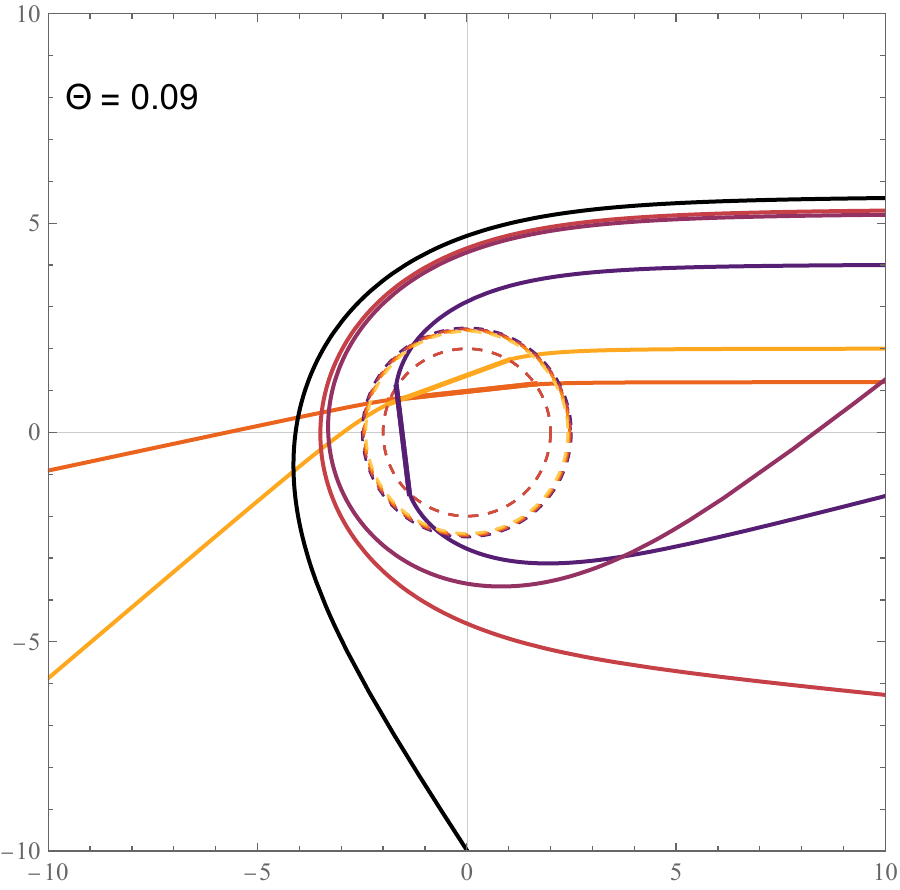}}
\subfigure[]{\includegraphics[scale=0.25]{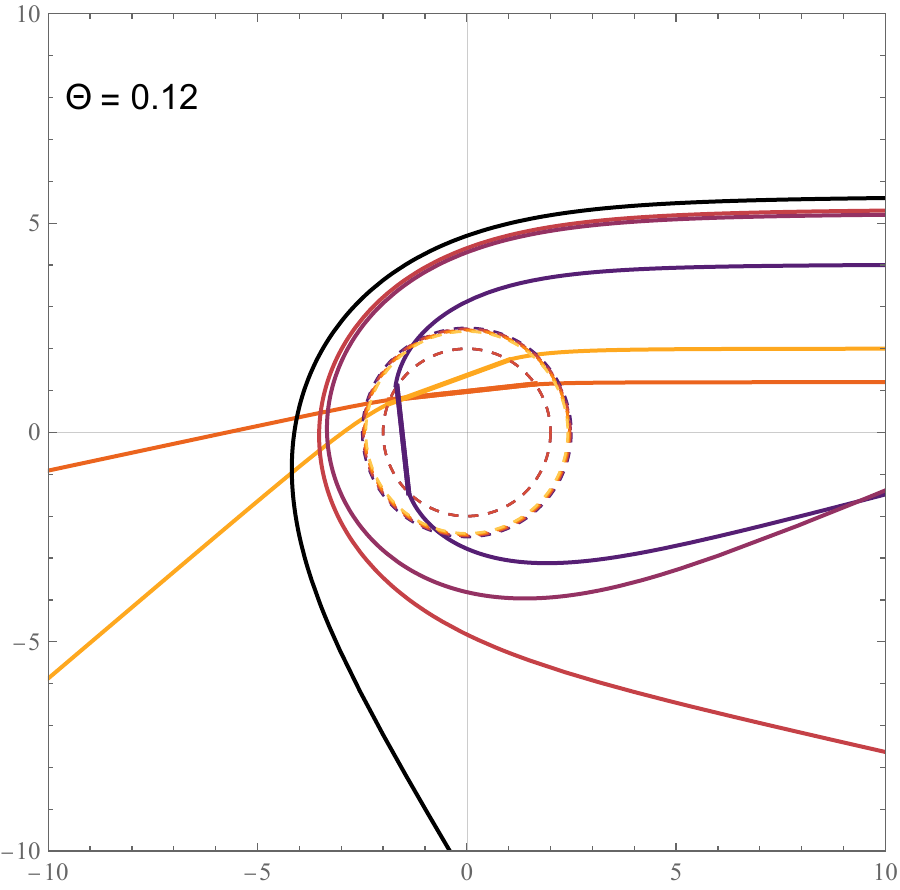}}
\caption{Influence of the noncommutative parameter on photon trajectories. Panels correspond to (a) $\Theta = 0.03$, (b) $\Theta = 0.06$, (c) $\Theta = 0.09$ and (d) $\Theta = 0.12$, while the same set of impact parameters $(b=1.2,2,4,5.2,5.3,5.6)$ is maintained in every panel. The selected values of $\Theta$ demonstrate that increasing it systematically modifies the photon deflection angle and shifts the critical impact parameter.}
\label{geod}
\end{figure}

We present the null geodesic motion in Fig.~\ref{Veff}, considering different values for the impact parameter $b$, which is associated with the energy and angular momentum. In the analysis, we fix the radius of the transparent shell at $\tilde{a}_0 = 2$, indicated in the figure as the inner circle, while $r_s = 2.5$ marks the boundary where the outer spacetime is described by the noncommutative Schwarzschild metric. Unlike the case of a black hole, even if a photon crosses the unstable circular orbit, $r_s = 2.5$, it eventually passes through the gravastar and scatters.

For the external region of the gravastar, using  (\ref{geo42}) and introducing a new variable $u = 1/r_{+}$, we have \cite{Campos:2021sff}
\begin{equation}
\dfrac{d u}{d\varphi} = \sqrt{\dfrac{1}{b^2} - \dfrac{1}{r^2} + \dfrac{2}{r^3} - \dfrac{\Theta}{~r^4}},
\label{eqD1.1}
\end{equation}
where $b = L/E$ is the impact parameter defined as the perpendicular distance (measured at infinity) between the geodesic and a parallel line that goes to the shell.

Differentiating \eqref{eqD1.1}, we obtain
\begin{equation}
\dfrac{d^{2}u}{d\varphi^{2}} = \dfrac{1}{r^3} - \dfrac{3}{r^{4}} + \dfrac{2\Theta}{~r^{5}}.
\label{eqD2.2}
\end{equation}
In this case, we can find a critical radius or critical circular orbit for a photon, $r_c$ and $b_c$, respectively, by using the conditions: $V(r_c) = 0$ and ${d V(r_c)}/{dr}= 0$. These conditions yield
\begin{equation}
r_c = \dfrac{3  + \sqrt{9 - 8~\Theta}}{2},
\label{r_c}
\end{equation}
\begin{equation}
b_c = \dfrac{r_c}{\sqrt{f(r_c)}} = \dfrac{r_c}{\sqrt{1-\dfrac{2}{r_c}+\dfrac{\Theta}{~r_{c}^{2}}}}.
\label{b_c}
\end{equation}

Fig.~(\ref{Veff}) shows that the impact parameter $b$ determines the proximity and deflection of photons: smaller values of $b$ lead to trajectories closer to the inner region, with steeper curves, while larger values of $b$ result in smaller deviations, typical of weak scattering. Using equations (\ref{r_c}) and (\ref{b_c}) we obtain $b_c\approx 5.143$. When we compare our results with those obtained in \cite{Sakai:2014pga}, we realize that both exhibit similar behavior.

Then when we change the value of the noncommutativity parameter $\Theta$, the differences between the curves mainly reflect the variation of $b$, which allows a comparison of the behavior of this parameter, as shown in Fig.~\ref{geod}. When comparing the results obtained in \cite{Campos:2021sff}, we observed that the results were analogous regarding the behavior of the external region of the gravastar, differing only in the internal region, since this UCO does not present a singularity. Thus, the photons pass through the object instead of being absorbed, because the shell is transparent in our study.
To examine more clearly the radius that defines the shadow size and the effect of the non-commutative parameter on it in the external region of the gravastar, we can use the results presented in ~\cite{Campos:2021sff}.

{In contrast to a Schwarzschild black hole, a gravastar lacks an event horizon. In the model addressed in this study, the thin shell is considered transparent. Consequently, incident photons are not permanently absorbed; instead, they can traverse the interior spacetime and re-emerge toward a distant observer. However, the geometry allows for unstable circular photon orbits determined by the critical radius $r_c$, which defines a critical impact parameter $b_c$ separating distinct classes of null geodesics. Photons with $b \simeq b_c$ undergo strong deflections and may orbit the compact object multiple times before escaping to infinity, which considerably reduces the radiation flux from these directions and creates a shadow-like optical structure, even though it arises solely from gravity and is not linked to capture by an event horizon \cite{Sakai:2014pga,Rosa:2024bqv}. Therefore, in this study, the term \textit{shadow} refers to this region of photon depletion created by the gravitational lensing structure of the spacetime.}

{This differentiation leads to an observational signature that is qualitatively distinct from the one produced by a black hole. In the Schwarzschild case, photons with $b < b_c$ are absorbed by the event horizon, resulting in a dark central region,  as shown in the study by Campos~\cite{Campos:2021sff}. In a transparent gravastar, these photons traverse the cascade, propagate through the interior of the Sitter, and can be detected by the observer as they emerge from the opposite side. Consequently, the predicted image is a dark ring around a bright central region, composed of photons traversing the interior of the object \cite{Sakai:2014pga}.}
\begin{eqnarray}
\alpha_{\rm sh}=\arcsin\left(\frac{b_c}{D}\right)\simeq\frac{b_c}{D},\qquad (D\gg b_c),
\end{eqnarray}
{Thus, any alteration in the photon sphere caused by the noncommutative parameter $\Theta$ directly modifies the critical impact parameter $b_c$ and, consequently, the apparent angular radius of the shadow. Since the photon trajectories are also governed by $\Theta$, the corresponding gravitational deflection angle is likewise affected, as qualitatively illustrated in Figs.~(\ref{Veff}) and (\ref{geod}), where increasing $\Theta$ systematically changes the bending of null geodesics. Therefore, both the angular size of the shadow-like feature and the associated lensing pattern constitute direct optical observables that may distinguish transparent gravastars from Schwarzschild black holes and provide signatures of noncommutative geometry.}

\section{STABILITY OF THE GRAVASTAR}\label{STA}
Now, we are interested in verifying the stability of the gravastars {in noncommutative geometry. In order to achieve this, we utilize the surface energy density $\sigma(a)$ on the gravastars' thin shell as follows\cite{Lobo:2015lbc}:}
\begin{eqnarray}
\dfrac{1}{2}\dot{a}^2 + V(a) = 0,
\end{eqnarray}
where the potential,
\begin{eqnarray}
    V(a)= \dfrac{1}{2} \left[1 - \dfrac{\mathcal{G}_1 (a)}{a} - \left( \dfrac{m_S (a)}{2a}\right)^{2} - \left( \dfrac{\mathcal{G}_2 (a)}{m_S(a)}\right)^{2} \right]. 
\end{eqnarray}
{The quantities $\mathcal{G}_1$ and $\mathcal{G}_2$, for simplicity, are defined as}
\begin{eqnarray}
    \mathcal{G}_1(a) = \dfrac{1}{2}\left[\left(2M +\dfrac{\Theta M}{\tilde{a}}\right) - a \tilde{\Lambda}\right],
\end{eqnarray}\begin{eqnarray}
    \mathcal{G}_2(a) = \dfrac{1}{2}\left[\left(2M +\dfrac{\Theta M}{\tilde{a}}\right) + a \tilde{\Lambda}\right].
\end{eqnarray}
{It is important to note that the effective potential $V(a)$ depends explicitly on the surface mass $m_s(a)$. In certain situations, it may be advantageous to invert this relationship and determine the surface mass as a function of the potential, rather than following the conventional procedure:}

\begin{eqnarray}
    m_s(a)= -a\left[\sqrt{1 - \frac{2 }{\tilde{a}} + \frac{\Theta}{\Tilde{a}^{2}} - 2V(a)}\;-\sqrt{1+ \tilde{\Lambda} - 2V(a)}\;\right]
\end{eqnarray}
{To investigate the stability of static configurations under radial perturbations, we consider a linearization around an equilibrium radius $a_0$ and perform a second-order Taylor expansion of the effective potential $V(a)$ about this point.} 
\begin{eqnarray}
    V(a)=V(a_0) + V'(a_0)(a - a_0) +\frac{1}{2}V''(a_0)(a-a_0)^2+\mathcal{O}\left[(a-a_0)^3\right],
\label{eq:taylor1}
\end{eqnarray}
{where the prime denotes the derivative with respect to $a$. For a static configuration, characterized by $\dot{a}_0=\ddot{a}_0=0$, the equilibrium conditions imply that $V(a_0)=V'(a_0)=0$. }
\begin{eqnarray}
    V(a)=\frac{1}{2}V''(a_0)(a-a_0)^2+\mathcal{O}\left[(a-a_0)^3\right],
\label{eq:taylor}
\end{eqnarray}
{Therefore, to evaluate a static equilibrium configuration at $a_0$, is stable if and only if $V(a)$ has a minimum which requires $V''(a_0)>0$.}

In this context, the response of the thin shell under radial perturbations can be characterized by the parameter $\eta$, which can be defined as the ratio between the surface pressure differential $\Tilde{\mathscr{P}}$ and surface density $\Tilde{\sigma}$, determined on the hypersurface $r = a_0$. In symbolic terms, as discussed in \cite{BHAR2021}. In dimensionless form, we have
\begin{equation}
\eta = \dfrac{\Tilde{\mathscr{P}}'}{\Tilde{\sigma}'}
\label{t4}
\end{equation}
where the prime denotes differentiation with respect to the shell
radius.

{It is important to emphasize that, although the parameter $\eta$ is frequently interpreted as the squared speed of sound in ordinary continuous media, such an interpretation is not strictly applicable to an infinitely thin surface layer described by the Darmois-Israel formalism. In the present context, $\eta$ should therefore be regarded as an effective response parameter relating variations of the surface pressure and surface energy density. Consequently, values $\eta>1$ do not, by themselves, imply superluminal signal propagation or causality violation.}

{The stability criterion is determined through the effective-potential condition $V''(a_0)>0$. In Fig.~(\ref{G6a}), the parameter $\eta$ is shown as a function of the dimensionless shell radius for different values of $\Theta$. The region denoted by ``S'' represents the range of configurations associated with the stable sector of the linearized thin-shell analysis. Therefore, $\eta$ is used here as an effective response parameter characterizing the shell within the stability
region, rather than as an independent stability criterion.}

\begin{figure}[h!]	
\centering 
\includegraphics[scale=.3]{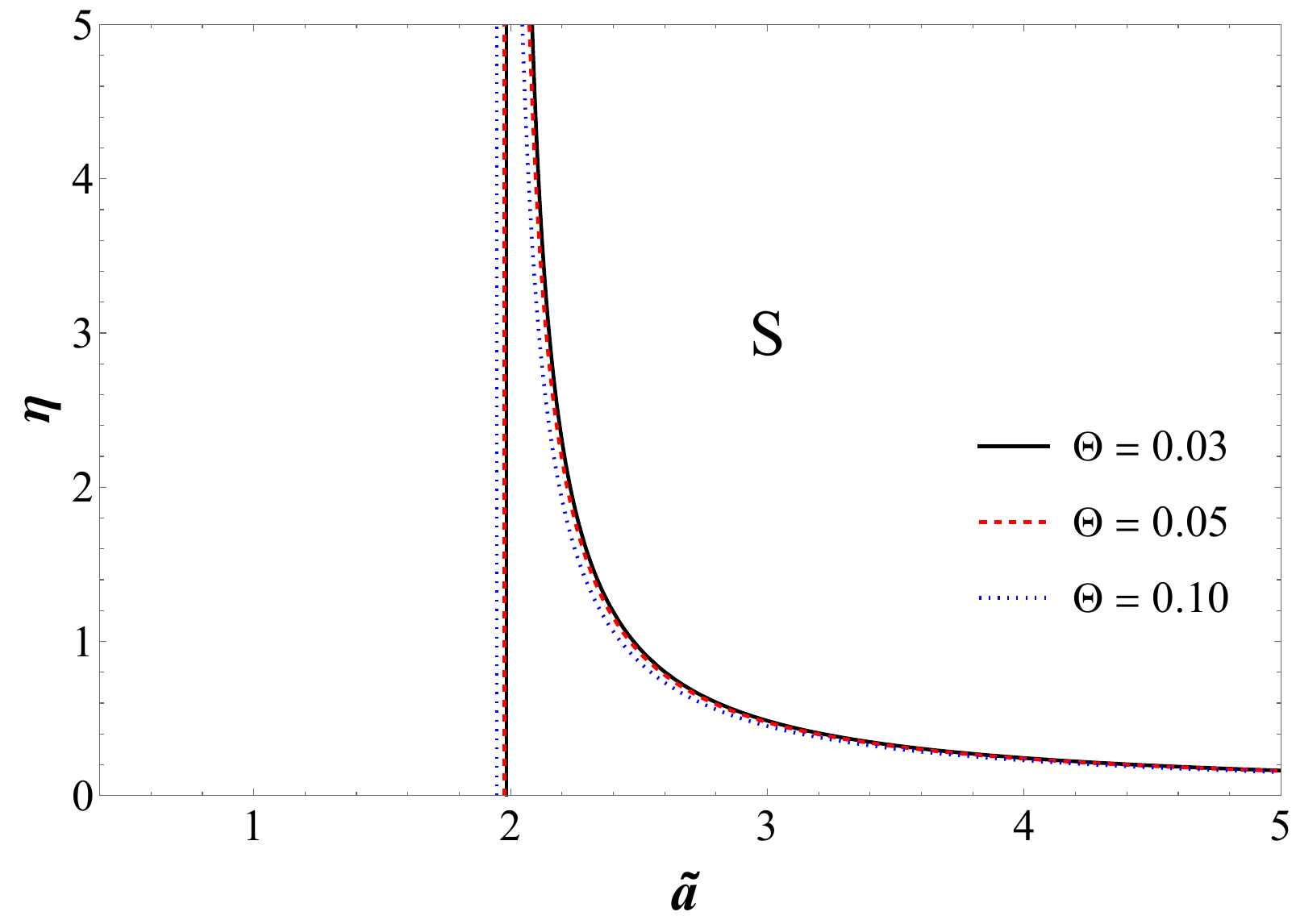} 
\caption{Effective response parameter $\eta$ as a function of the dimensionless shell radius $\tilde a$ for $\Theta=0.03$, $0.05$, and $0.10$. The region denoted by ``S'' indicates the stable sector considered in the linearized thin-shell analysis. The weak dependence of the curves on $\Theta$ shows that the noncommutative corrections produce only a moderate modification of the shell response.}
\label{G6a}
\end{figure} 

\section{Conclusions}\label{sec4}

{Observationally identifying black holes remains an open problem in astrophysics. The central difficulty is that the formation of shadows, long treated as an exclusive signature of black holes, arises from the existence of unstable circular photon orbits, a property that other ultracompact objects may also possess. Gravastars provide a concrete example of this. It is this observational ambiguity that motivates their investigation.}

{We construct a thin-shell gravastar model incorporating noncommutative geometry effects in the exterior region, described by a Schwarzschild metric modified by a Lorentzian-type mass distribution. The interior is characterized by a de Sitter geometry, and the two regions are connected by a thin shell according to the Israel junction conditions. This construction captures, in a controlled manner, possible corrections arising from quantum gravity without sacrificing the analytical tractability of the model.}

{The verification of the energy conditions on the shell produced a nontrivial result. For sufficiently small values of the noncommutative parameter $\Theta$, the weak, null, and strong energy conditions are all satisfied, even when the effective cosmological constant is zero, $\tilde{\Lambda} = 0$. In this regime, $\Theta$ assumes the role of stabilizing the shell, a function that would normally be attributed to vacuum energy. Noncommutativity of spacetime, therefore, is not merely a formal correction: it plays a structural role in the stability of the object.}

{The relation we establish between $\Theta$ and the cosmological constant, applied to a ten-solar-mass black hole, leads to a noncommutative energy scale of the order of $10~\mathrm{TeV}$. Quantum gravity effects are usually placed at the Planck scale, which is experimentally inaccessible by any reasonable prospect. An estimate of $10~\mathrm{TeV}$ obtained from geometrical and astrophysical considerations is therefore an unexpected result: it places these effects in a regime that next-generation accelerators may, at least in principle, be able to probe.}

{In comparison with the work of Sakai et al.~\cite{Sakai:2014pga}, which pioneered the investigation of gravastar ``shadows,'' our model introduces noncommutative corrections and shows how they affect light trajectories. The connection between $\Theta$ and the $10~\mathrm{TeV}$ scale introduces a link with particle physics that is absent in purely geometrical models. The noncommutative gravastar thus carries a well-defined physical energy signature.}

{The numerical integration of the null geodesic equations revealed a qualitative difference relative to the black hole case: photons with impact parameter below the critical value $b_c$ are not absorbed. They cross the transparent shell and emerge on the opposite side, producing luminous structures where a black hole would project darkness. This behavior is consistent with Sakai et al.~\cite{Sakai:2014pga}. 
{Fig.~(\ref{geod}) shows that increasing the noncommutative parameter $\Theta$ systematically modifies photon deflection, the critical impact parameter, and consequently the apparent angular radius of the shadow-like feature. These quantities constitute direct optical observables that may allow transparent gravastars to be distinguished observationally from Schwarzschild black holes.}

{The stability analysis, via the effective response parameter $\eta=\tilde{P}'/\tilde{\sigma}'$, identified a well-defined stability region, indicated by `S'' in Fig.~(\ref{G6a}). That $\eta$ may exceed unity in this region is physically admissible for thin shells, as discussed by Poisson and Visser~\cite{POISSON1995} and Lobo~\cite{Lobo:2003xd}. Our results are consistent with those reported in analogous contexts~\cite{Yousaf:2019zcb, Ovgun:2017jzt, Sharif:2021zzr}.}

{Some limitations of the present work directly point to future extensions. Spherical symmetry and the static regime are useful idealizations, but a more realistic treatment will need to incorporate emission models for the surface and interior of the gravastar. Nonlinear stability, response to non-spherical perturbations, emission spectra, and gravitational-wave signatures are relevant observables for next-generation detectors.}

{A thin-shell gravastar with a noncommutative exterior geometry is a physically consistent alternative to the classical black hole. Its geometric structure carries implications that connect observational astrophysics and particle physics, and this alone justifies further investigation of the model.}

{Future investigations could expand upon the current analysis by employing a full ray-tracing treatment of photon propagation through the transparent shell. This would enable the generation of image-plane maps, intensity distributions, and quantitative gravitational lensing observables. Specifically, calculating deflection angles in both weak and strong lensing regimes, combined with detailed brightness profiles and radiative transfer effects, would allow for a more direct link between the noncommutative parameter $\Theta$ and astrophysical observations. Such analyses could establish more robust observational criteria for distinguishing transparent noncommutative gravastars from classical black holes, while also clarifying the optical signatures associated with horizonless compact objects.}
\begin{acknowledgments}
We thank CNPq and CAPES for partial financial support. M.A.A  acknowledges support from CNPq (Grant no. $301683/2025-5$). A.T.N.S thanks the Paraíba State Research Support Foundation (FAPESQ) (Grant No. $765/2026$) for financial support. 
\end{acknowledgments}

\end{document}